%% file: main.tex
\documentclass[10pt, aps, prb, twocolumn, superscriptaddress, nobibnotes, floatfix]{revtex4-2}

\usepackage{amsmath, amsfonts}
\usepackage{physics}

\usepackage{graphicx}
\usepackage[colorlinks,allcolors=blue]{hyperref}
\usepackage{float}
\usepackage{url}

\begin{document}

\title{Nonperturbative decay of bipartite discrete time crystals}

\author{Lennart Fernandes}
\email{lf2680@nyu.edu}
\affiliation{Center for Quantum Phenomena, Department of Physics, New York University, New York, NY 10003, USA}

\author{Joseph Tindall}
\affiliation{Center for Computational Quantum Physics,
Flatiron Institute, New York, NY 10010, USA}

\author{Dries Sels}
\affiliation{Center for Quantum Phenomena, Department of Physics, New York University, New York, NY 10003, USA}
\affiliation{Center for Computational Quantum Physics,
Flatiron Institute, New York, NY 10010, USA}

\begin{abstract}
We study prethermal time-crystalline order in periodically driven quantum Ising models on disorder-free decorated lattices.
Using a tensor network ansatz for the state which reflects the geometry of a unit cell of the lattice, we show through finite entanglement scaling that the system has an exponentially long-lived subharmonic response in the thermodynamic limit, which decays nonperturbatively in deviations from a perfect periodic drive.
The resulting prethermal discrete time crystal is not only stable to imperfections in the transverse field, but also exhibits a bipartite rigidity to generic perturbations in the longitudinal field.
We call this state a bipartite discrete time crystal and reveal a rich prethermal phase diagram, including multiple regions of bipartite time-crystalline order, uniform time-crystalline order and thermalization, with boundaries depending delicately on the topology of the decorated lattice. Our results thus uncover a variety of time crystals which may be realized on current digital quantum processors and analog quantum simulators.
\end{abstract}

\maketitle 

\textbf{\emph{Introduction.}}~---~
The stable crystalline structure of solid materials is a striking example of spontaneous symmetry breaking, as the ground state violates spatial translational invariance in the formation of a periodic lattice. 
The temporal counterpart of this phenomenon, a state which evolves periodically in time, constitutes a breaking of time translational invariance which has been coined a \emph{time crystal}~\cite{wilczekQuantumTimeCrystals2012, sachaTimeCrystalsReview2018}.
While the spontaneous breaking of a continuous time symmetry by an isolated quantum system in equilibrium was proven to be impossible~\cite{brunoImpossibilitySpontaneouslyRotating2013, nozieresTimeCrystalsCan2013, watanabeAbsenceQuantumTime2015}, out-of-equilibrium states which break a discrete time translational invariance can be realized in Floquet systems, where an external drive modulates the governing Hamiltonian over a period $\tau$, $\mathcal{H}(t) = \mathcal{H}(t+\tau)$~\cite{floquetEquationsDifferentiellesLineaires1883, shirleySolutionSchrodingerEquation1965, moessnerEquilibrationOrderQuantum2017}.
In this case, the discrete time symmetry imposed by the periodic drive is broken when an observable $O$ does not stabilize at stroboscopic times $t=n\tau$, but instead exhibits long-lived periodicity at a lower frequency than that of the Hamiltonian,
\begin{equation}
    \langle O(t)\rangle = \langle O(t+\tilde{n}\tau)\rangle, \quad (\tilde{n}>1).
\end{equation}

When stable to generic local perturbations, such symmetry-broken states are called \emph{discrete time crystals}~(DTC)~\cite{sachaModelingSpontaneousBreaking2015, elseDiscreteTimeCrystals2020, zaletelColloquiumQuantumClassical2023} due to the spontaneous emergence of a subharmonic order parameter $\langle O \rangle$.
The primary obstacle for constructing long-lived
DTCs is the absorption of energy injected by the Floquet drive, which causes the system to relax to an infinite temperature equilibrium state.
With the exception of nonergodic dynamics in quantum-scarred eigenstates~\cite{turnerWeakErgodicityBreaking2018, bluvsteinControllingQuantumManybody2021, maskaraDiscreteTimeCrystallineOrder2021, huangDiscreteTimeCrystals2022} and systems exhibiting many-body localization due to disorder~\cite{nandkishoreManyBodyLocalizationThermalization2015, abaninColloquiumManybodyLocalization2019, elseFloquetTimeCrystals2016,  khemaniPhaseStructureDriven2016, yaoDiscreteTimeCrystals2017, zhangObservationDiscreteTime2017,
choiObservationDiscreteTimecrystalline2017, freyRealizationDiscreteTime2022}, linear external fields~\cite{schulzStarkManyBodyLocalization2019, vannieuwenburgBlochOscillationsManybody2019, kshetrimayumStarkTimeCrystals2020, liuDiscreteTimeCrystal2023} or gauge symmetries~\cite{smithDisorderFreeLocalization2017, brenesManyBodyLocalizationDynamics2018, russomannoHomogeneousFloquetTime2020}, discrete time crystals in non-integrable many-body quantum systems are a transient phenomenon with a finite \emph{prethermal}~\cite{bukovPrethermalFloquetSteady2015, weidingerFloquetPrethermalizationRegimes2017, rubio-abadalFloquetPrethermalizationBoseHubbard2020, fleckensteinThermalizationPrethermalizationPeriodically2021, hoQuantumClassicalFloquet2023} lifetime.

Prethermal time-crystalline order was notably shown to have a lifetime scaling exponentially with the driving frequency~\cite{abaninExponentiallySlowHeating2015, moriRigorousBoundEnergy2016, abaninEffectiveHamiltoniansPrethermalization2017}, and has since been studied extensively in a wide variety of systems~\cite{elsePrethermalPhasesMatter2017, zengPrethermalTimeCrystals2017, luitzPrethermalizationTemperature2020, machadoLongRangePrethermalPhases2020, pizziHigherorderFractionalDiscrete2021, kyprianidisObservationPrethermalDiscrete2021, santiniCleanTwodimensionalFloquet2022, vuDissipativePrethermalDiscrete2023, beatrezCriticalPrethermalDiscrete2023, xiangLonglivedTopologicalTimecrystalline2024, shinjoUnveilingCleanTwodimensional2024, jiangPrethermalTimeCrystallineCorner2024, moonDiscreteTimeCrystal2024, shuklaPrethermalFloquetTime2024}.
Among the mechanisms proposed to extend DTCs beyond the limiting case $\tau \to 0$~\cite{zaletelColloquiumQuantumClassical2023, ikedaRobustEffectiveGround2024} is the presence of a confining potential, which, in analogy to the confinement of quarks in QCD~\cite{wilsonConfinementQuarks1974, sulejmanpasicConfinementBulkDeconfinement2017}, causes magnetic domain walls to form bound states as a result of the energy cost associated with the growth of the domain~\cite{lakeConfinementFractionalQuantum2010, jamesNonthermalStatesArising2019, ramosConfinementBoundStates2020}.
The consequently inhibited spread of correlations was found to delay the onset of thermalization~\cite{kormosRealtimeConfinementFollowing2017, birnkammerPrethermalizationOnedimensionalQuantum2022, tindallConfinementTransverseField2024}, and shown to allow for long-lived time-crystalline order in the case of confinement by an external field in a one-dimensional Ising chain~\cite{colluraDiscreteTimeCrystallineResponse2022}.
Since symmetry-broken states of ferromagnetic models in higher dimensions generically exhibit confinement due to the growth of a domain wall with the size of the domain itself ~\cite{tindallConfinementTransverseField2024, pavesicConstrainedDynamicsConfinement2024}, disorder-free DTCs have been predicted to occur in the two-dimensional square Ising model~\cite{santiniCleanTwodimensionalFloquet2022}, 
and were recently observed on a digital quantum processor with a decorated hexagonal geometry~\cite{shinjoUnveilingCleanTwodimensional2024}. 
In the latter system, both Floquet dynamics and non-thermal dynamics caused by confinement of excitations have been accurately reproduced through tensor network state (TNS) simulations~\cite{tindallEfficientTensorNetwork2024, tindallConfinementTransverseField2024}.

\, 

We build upon these successes to study the emergence, stability and lifetime of time-crystalline order in the Floquet Ising model on generic decorated lattice structures. By employing a tensor network reflecting the unit cell of an infinite realization of the lattice, optimized with belief propagation, we directly access the thermodynamic limit. 
Supported by state vector simulations on finite graphs, this will lead us to introduce \emph{bipartite} discrete time crystals and construct a prethermal phase diagram hosting several DTC regimes determined by the lattice topology.

\begin{figure}[tbp]
    \centering
    \includegraphics[width=\columnwidth]{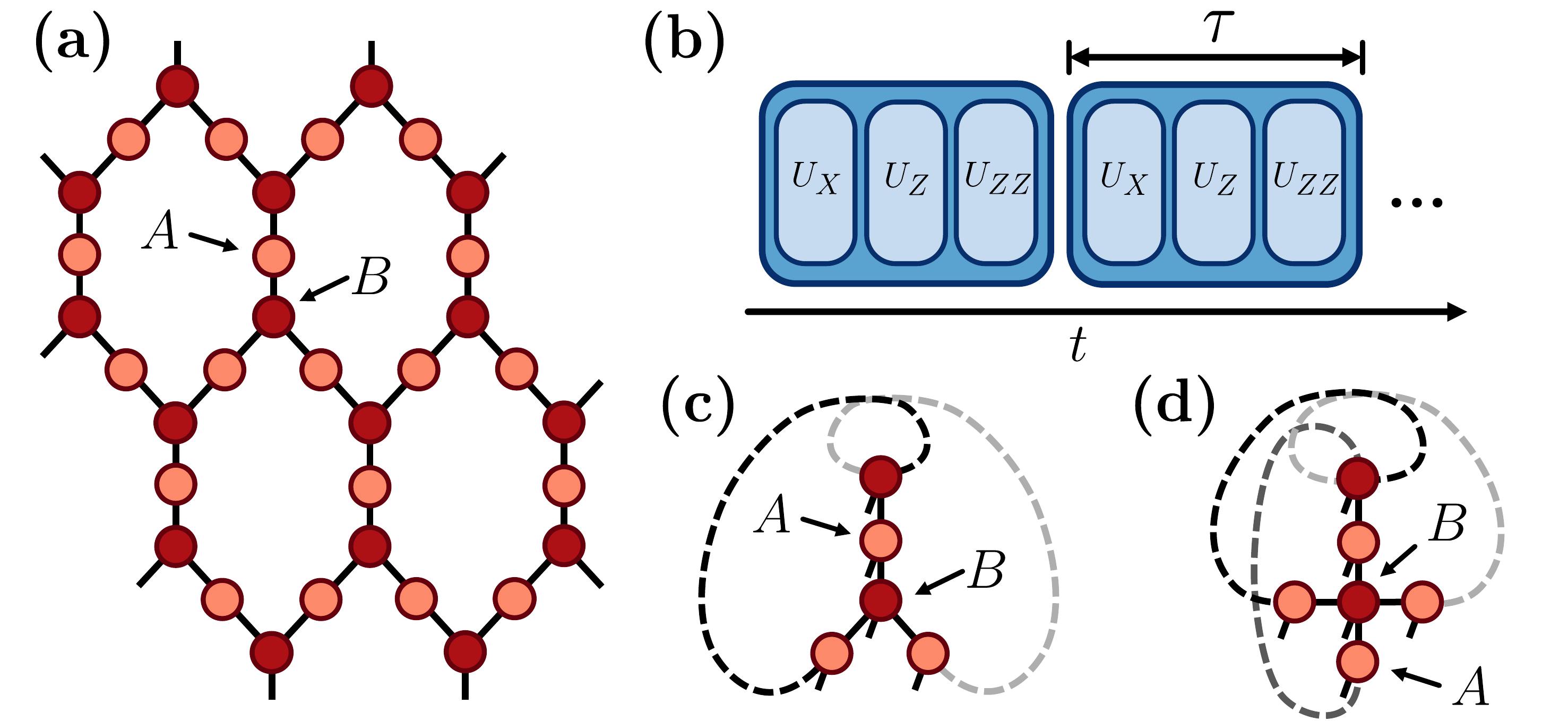}
    \caption{\textbf{Geometry and dynamics.} (a)~The two-dimensional decorated hexagonal lattice consists of two sublattices of sites with two ($A$, light) and three ($B$, dark) nearest neighbors. (b)~Unitary sequence of two Floquet cycles. (c)~The complete bipartite graph $K_{2,3}$ as the BP-iTNS unit cell of a decorated hexagonal lattice, with periodic boundaries (dashed lines). (d)~The $K_{2,4}$ unit cell of a Lieb lattice.}
    \label{fig:system}
\end{figure}

\vfill 
\pagebreak
\textbf{\emph{Model and approach.}}~---~
We consider the Ising model in the presence of both transverse and longitudinal magnetic fields, described by the Hamiltonian
\begin{equation}
    \mathcal{H} = h_x \sum_j X_j + h_z \sum_j Z_j  - J \sum_{\langle jk \rangle} Z_j Z_k,
    \label{eq:hamiltonian}
\end{equation}
where $X$ and $Z$ are the spin-$1/2$ Pauli operators. Both the interaction strength $J$ and the fields $h_x$ and $h_z$ are taken to be uniform, while the summations $j$ and $\langle jk \rangle$ run over all sites and nearest-neighbor links of the underlying decorated lattice. We define a decorated lattice as one in which each of the edges of a lattice with vertices of uniform coordination number $z$ is decorated with an additional vertex, with the example of the decorated hexagonal lattice shown in Fig.~\ref{fig:system}(a).
A distinguishing feature of this geometry is its bipartite structure, comprising two distinct sublattices of vertices with degree $2$ and $z > 2$. We will refer to these sublattices as the $A$- and $B$-lattices, respectively.
A discrete time symmetry is introduced by the periodically driven dynamics illustrated in Fig.~\ref{fig:system}(b), in which unitary time evolution over the period $\tau$ of a single Floquet cycle is generated by successive application of the three contributions to the Hamiltonian for a time $\tau/3$. 
Setting $\hbar=1$ and defining the rotation operator of an operator $O$ as $R_O(\theta) = e^{-i\theta O/2}$, we can write this Floquet unitary as
\begin{equation}
    U(\tau) = \bigg[\prod_j R_{X}^{j}(\theta_x)\bigg]\bigg[\prod_j R_{Z}^{j}(\theta_z)\bigg]\bigg[\prod_{\langle jk \rangle} R_{Z Z}^{jk}(\theta_J)\bigg], 
    \label{eq:floquet_unitary}
\end{equation}
where the rotation angles are related to the Hamiltonian and the Floquet period $\tau$ as $\theta_J = -2J \tau/3$, $\theta_x=2h_x \tau/3$ and $\theta_z=2h_z \tau/3$.

To compute the dynamics induced by $U(\tau)$ on a given decorated lattice in the thermodynamic limit, we use the BP-iTNS ansatz \cite{tindallGaugingTensorNetworks2023, tindallEfficientTensorNetwork2024, tindallConfinementTransverseField2024}. Specifically, we simulate the dynamics induced by the propagator in Eq.~\eqref{eq:floquet_unitary} on the $K_{2,z}$ complete bipartite graph \cite{bollobasModernGraphTheory1998}. This geometry corresponds to a single unit cell of the decorated lattice with periodic boundaries. We use a tensor network state (TNS) whose structure matches that of the $K_{2,z}$ graph. The application of two-site gates and the measurement of expectation values is done via belief propagation (BP), which works by identifying vectorized message tensors as a rank-$1$ approximation for the environments in the tensor network~\cite{leiferQuantumGraphicalModels2008,alkabetzTensorNetworksContraction2021,sahuEfficientTensorNetwork2022,guoBlockBeliefPropagation2023,tindallEfficientTensorNetwork2024}. These BP-iTNS results are exactly equivalent \cite{tindallGaugingTensorNetworks2023,tindallConfinementTransverseField2024, tindallEfficientTensorNetwork2024} to those obtained by simulating the dynamics of the full decorated lattice in the thermodynamic limit  with a TNS matching the infinite lattice and optimized with BP.
The error made by the BP approximation decreases exponentially with the size of the smallest loop in the system, making it especially suitable for the considered geometries due to the fact the decorations double the smallest loop size in the original lattice. 
Within the limitations set by the bond dimension of the iTNS and the loop corrections neglected by the BP approximation, the BP-iTNS ansatz allows us to access the dynamics $U(\tau)$ on any decorated lattice structure in the thermodynamic limit. We picture this ansatz for the decorated hexagonal ($z=3$) and decorated square (Lieb) lattice ($z=4$) in Fig.~\ref{fig:system}(c-d).

\,

\textbf{\emph{Time-crystalline order.}}~---~
We consider the system initialized in the symmetry-broken product state $\ket{\uparrow}^{\otimes (z+2)}$ and subject to an effective confining potential through interactions, taken to be maximally entangling two-qubit rotations $\theta_J=-\pi/2$ at a vanishing longitudinal field $\theta_z=0$.
For a transverse field $\theta_x=\pi$, the dynamics consist of $R_{ZZ}$ interactions alternating with a $\pi$-flip of all qubits around the $X$-axis, resulting in a perfect time-crystalline order of the magnetization with a subharmonic period $2\tau$,
\begin{equation}
    \langle \psi (n\tau)|Z|\psi (n\tau) \rangle = (-1)^n.
\end{equation}
For any imperfect drive $\theta_x\neq \pi$, the system leaves the $Z$-polarized state and eventually thermalizes. 
However, as the stroboscopic dynamics in the limit $\theta_x \to \pi$ are governed to lowest order by a static Ising Hamiltonian, confinement dramatically slows down the absorption of energy and generation of entanglement~\cite{tindallConfinementTransverseField2024}, delaying thermalization during a prethermal time-crystalline stage. 

\begin{figure}[tbp]
    \centering
    \includegraphics[scale=1]{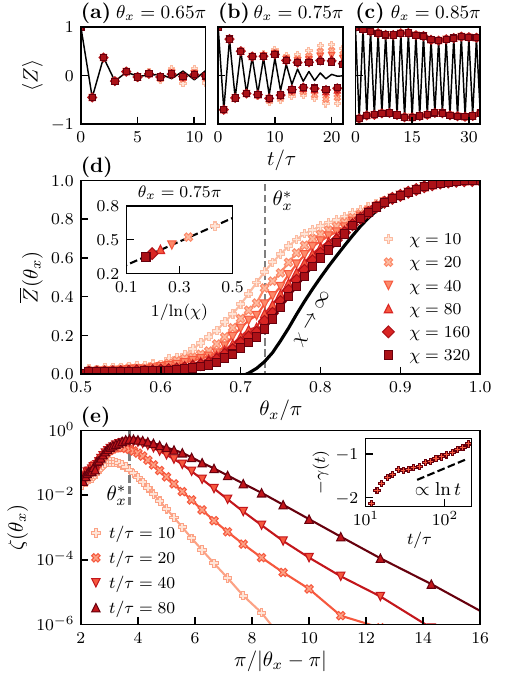}
    \caption{\textbf{Time-crystalline order.} (a-c)~Evolution of the magnetization on an infinite decorated hexagonal lattice for a periodic transverse field drive in the thermalizing (a), crossover (b) and DTC regime (c), evaluated at the bond dimensions shown in (d). (d)~Mean magnetization in the thermodynamic limit as a function of $\theta_x$, averaged over the first $100$ cycles at different bond dimensions. The inset illustrates for $\theta_x=0.75\pi$ the extrapolation $1/\ln{(\chi)} \to 0$, resulting in the full black line in panels (a-d). The crossover value $\theta_x^{*}$ inferred from a finite entanglement scaling is shown by the grey dashed line in (d). (e)~Slope of the extrapolation \eqref{eq:Z_extrapolation} for different number of Floquet cycles, exhibiting exponential decay with $1/|\theta_x-\pi|$ in the DTC regime. The inset shows the evolution of the exponent defined in Eq. \eqref{eq:exp_decay}.}
    \label{fig:DTC_order}
\end{figure}

Figure~\ref{fig:DTC_order}(a-c) shows the time evolution of the magnetization $\langle Z \rangle$ in the $A$-sublattice of a decorated hexagonal lattice subject to Floquet dynamics at three different transverse field drives $\theta_x/\pi=\{0.65, 0.75, 0.85 \}$. For rotation angles deviating significantly from the perfect $\pi-$periodic drive as in Fig.~\ref{fig:DTC_order}(a), $\langle Z \rangle$ dephases rapidly as the system relaxes to an infinite temperature state. Conversely, Fig.~\ref{fig:DTC_order}(c) reveals a long-lived subharmonic response in the limit $\theta_x \to \pi$, signaling the emergence of a discrete time crystal. 
A crossover between these dynamical regimes of DTC order and thermalization was recently observed in error-mitigated quantum simulations on IBM's Heron architecture~\cite{shinjoUnveilingCleanTwodimensional2024}. 
We develop a more exhaustive picture of this crossover directly in the thermodynamic limit by plotting in Fig.~\ref{fig:DTC_order}(d) the mean oscillation amplitude $\overline{Z}(\theta_x, \chi)$, defined as the average of $|\langle Z(n\tau) \rangle|$ during the first $t/\tau = 100$ cycles of the Floquet unitary \eqref{eq:floquet_unitary}.
In agreement with the dynamics illustrated in Fig.~\ref{fig:DTC_order}(a-c), the amplitude reaches zero as $\theta_x$ approaches $\pi/2$, and converges to unity in the limit $\theta_x \to \pi$. In the latter regime, the insensitivity of the magnetization to the bond dimension $\chi$ of the employed BP-iTNS implies that entanglement growth is suppressed, consistent with the presence of an effective confining potential.
For intermediary values of $\theta_x$, a slow convergence with respect to $\chi$ indicates that the system becomes highly entangled as it crosses over into the thermalizing regime.

The scaling of the mean magnetization $\overline{Z}$ with the bond dimension allows us to predict the time-crystalline dynamics at an unbounded entanglement entropy $S_E \propto \ln(\chi)$ by extrapolating our results at finite bond dimension to the limit $\ln(\chi) \to \infty$. As illustrated in the inset of Fig.~\ref{fig:DTC_order}(d), we find that $\overline{Z}(\theta_x, \chi)$ for any $\theta_x$ scales with the bond dimension as
\begin{equation}
    \overline{Z}(\theta_x,\chi)=\frac{\zeta(\theta_x)}{\ln(\chi)} + \overline{Z}_\infty(\theta_x).
    \label{eq:Z_extrapolation}
\end{equation}
On the condition that the slope $\zeta(\theta_x)$ of this entanglement scaling is finite, the quantity $\overline{Z}_\infty(\theta_x)$ corresponds to the prethermal magnetization in the limit $\chi \to \infty$, indicated by the full black lines in Fig.~\ref{fig:DTC_order}(a-d).
The slope $\zeta(\theta_x)$ signifies the dependence of $\overline{Z}$ on the bond dimension, and hence serves as a measure of entanglement production during the dynamics. This quantity is plotted in Fig.~\ref{fig:DTC_order}(e) as a function of the inverse perturbation $\pi/|\theta_x-\pi|$ to a $\pi$-periodic drive, for $\overline{Z}$ evaluated over a varying number $t/\tau$ of Floquet cycles. From this relation, the growth of entanglement is found to reach a maximum at a drive field which converges at late times to $\theta_x^{*}/\pi\approx 0.73$, providing an estimate of the crossover between DTC and rapidly thermalizing dynamics in agreement with the range $\theta_x^{*}/\pi\sim 0.7 - 0.8$ inferred from quantum simulation~\cite{shinjoUnveilingCleanTwodimensional2024}.
Moreover, Fig.~\ref{fig:DTC_order}(e) reveals that the entanglement production in the DTC regime scales as 
\begin{equation}
   \zeta(\theta_x, t) \sim  \textnormal{exp}\{-\gamma(t) \pi /|\theta_x-\pi|\},
   \label{eq:exp_decay}
\end{equation}
indicating that the decay of time-crystalline order is nonperturbative in the deviation $|\theta_x - \pi|$ from a perfect $\pi$-periodic drive.
Finally, the time evolution of the exponent $\gamma(t)$ provides an estimate for the lifetime of the prethermal DTC. The drift of $\gamma(t)$ to zero in the limit $t \to \infty$, seen in Fig.~\ref{fig:DTC_order}(e), indicates that the production of entanglement entropy eventually results in thermalization for any imperfect drive $\theta_x < \pi$ at late times. Nevertheless, the inset shows that this decay of $\gamma(t)$ occurs at most logarithmically in time, suggesting a DTC lifetime which grows exponentially in $1/|\theta_x-\pi|$.
More precisely, the relation $-\gamma(t) = \ln (t/t^*)$ implies that the entanglement production \eqref{eq:exp_decay} within the DTC regime and up to a characteristic time scale $t^*$ is of the form
\begin{equation}
    \zeta(\theta_x, t)  \sim (t/t^*)^{\pi/| \theta_x - \pi |},
\end{equation}
which is significantly lower than the linear (volume law) growth of entanglement entropy characterizing rapid thermalization.

\textbf{\emph{Bipartite rigidity.}}~---~
A comparable decay of the order parameter and a crossover value $\theta_x^{*}/\pi \sim 0.6- 0.8$ were previously inferred through matrix product state (MPS) simulations on finite square lattices~\cite{santiniCleanTwodimensionalFloquet2022}, indicating that prethermal subharmonic oscillations are not specific to decorated lattice structures. 
However, true time-crystalline order requires, besides a robustness of the subharmonic response to an imperfect periodic drive, a self-stabilizing rigidity with respect to generic local perturbations~\cite{elseDiscreteTimeCrystals2020}.
We therefore extend the Floquet unitary \eqref{eq:floquet_unitary} to include a finite longitudinal field $\theta_z>0$. 
Where the interference of two driving fields will cause a generic system to exhibit a beating, a system in a stable time-crystalline phase retains its subharmonic response.

\begin{figure}[tbp]
    \centering
    \includegraphics[scale=1]{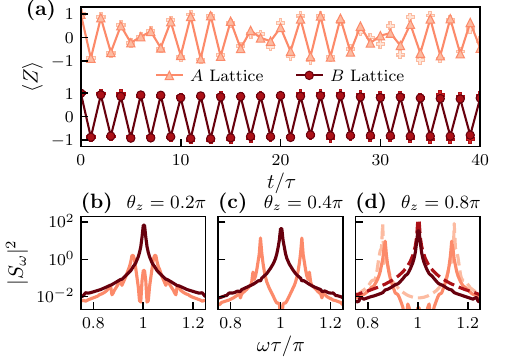}
    \caption{\textbf{Bipartite rigidity.} (a)~Time evolution of $\langle Z \rangle$ on the $A$- (top) and $B$-sublattice (bottom) of a decorated hexagonal lattice for $(\theta_J, \theta_x, \theta_z)=(-0.5\pi, 0.85\pi, 0.4\pi)$, calculated in the thermodynamic limit at $\chi=160$. (b-d)~Power spectra \eqref{eq:power_spectrum} of the magnetization in the first $t/\tau=400$ cycles, for $\theta_x=0.85\pi$ and $\theta_z/\pi=\{0.2, 0.4, 0.8\}$. The crosses in (a) and dashed lines in (d) indicate the corresponding result for $\chi=1$.}
    \label{fig:DTC_rigidity}
\end{figure}

Figure~\ref{fig:DTC_rigidity}(a) shows the response of $\langle Z \rangle$ on both sublattices of a decorated hexagonal lattice to a transverse drive of $\theta_x=0.85\pi$ within the time-crystalline regime, perturbed by a longitudinal field rotation $\theta_z=0.4\pi$. 
This configuration was reported in Ref.~\cite{shinjoUnveilingCleanTwodimensional2024} to cause a beating in the global magnetization, signalling the emergence of a discrete time quasi-crystal (DTQC)~\cite{ elseLongLivedInteractingPhases2020,auttiObservationTimeQuasicrystal2018,  heExperimentalRealizationDiscrete2024}. However, we find that the different connectivity of qubits in either sublattice results in qualitatively different dynamics. Whereas all qubits in the $A$-lattice exhibit a beating frequency manifesting as an envelope to the subharmonic response, a similar breaking of DTC order into quasi-crystalline dynamics is absent in the $B$-lattice, which is insensitive to longitudinal perturbations. This bipartite rigidity persists for all values of $\theta_z \in [0,\pi]$, as demonstrated in Figs.~\ref{fig:DTC_rigidity}(b-d) through the power spectrum of the magnetization dynamics,
\begin{equation}
    S_\omega = \frac{\tau}{t} \sum_{n=1}^{t/\tau} e^{-i n \omega \tau} \langle Z(n\tau) \rangle.
    \label{eq:power_spectrum}
\end{equation}
While the peak of the subharmonic response at $\omega \tau=\pi$ remains unperturbed by the longitudinal field in the $B$-sublattice, the instability in the $A$-lattice causes a splitting into two peaks separated by $\delta \omega \propto \theta_z$.
Since both sublattices exhibit reversible oscillatory dynamics, the entanglement entropy remains low.
As shown by the crosses in Fig.~\ref{fig:DTC_rigidity}(a) and dashed lines in Fig.~\ref{fig:DTC_rigidity}(b-d), this property allows us to observe all qualitative features even from the evolution of a product state ($\chi = 1$), suggesting that the bipartite response is not caused by intricate many-body dynamics, but is instead a necessary feature of the decorated lattice topology.

In the Supplemental Material~\footnote{See Supplemental Material for a convergence analysis of the obtained BP-iTNS results, additional details on the finite entanglement scaling and a comparison to exact state vector simulations of Floquet dynamics in finite graphs.}, we crosscheck the validity of our BP-iTNS method with exact state vector simulations of complete regular graphs with decorated edges, demonstrate the absence of ridigity in the corresponding undecorated lattices, and show that bipartite rigidity is not found in decorated square lattices ($z=4$).
Being a feature unique to systems with a bipartite topology, we thus define a \mbox{\emph{bipartite DTC}} as a nonequilibrium state of matter in which the inhibition of entanglement growth results in a robust subharmonic response in an extensive part of, but not all of, the system.

\, 

\begin{figure}[tbp]
    \centering
    \includegraphics[scale=1]{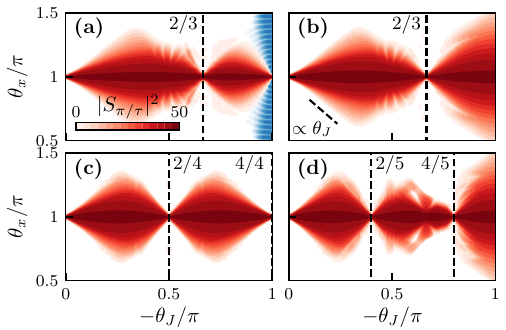}
    \caption{\textbf{Prethermal phase diagrams.} (a-b)~Spectral power density $|S_\omega|^2$ at the subharmonic peak $\omega \tau=\pi$ in the $A$- (a) and $B$-sublattice (b) of the Floquet Ising model ($\theta_z=0$) in the decorated hexagonal lattice, evaluated over 50 Floquet cycles in the limit $\chi\to \infty$. Red regions indicate the presence of DTC order, while the blue region in (a) represents the spectral density of the largest non-subharmonic peak, indicating quasi-crystalline enveloped oscillations. (c-d)~Subharmonic spectral power density $|S_{\pi/\tau}|^2$ in the $B$-sublattice of a Lieb lattice (c) and decorated 5-regular graph (d).}
    \label{fig:phasediagram}
\end{figure}

\textbf{\emph{Prethermal phase diagram.}}~---~
The decorated lattice topology not only causes a bipartite rigidity to perturbations, but determines the presence of time-crystalline order in the transverse field Ising model even at $\theta_z=0$.
To illustrate this, we extend the result of Fig.~\ref{fig:DTC_order}(d) to arbitrary interaction strengths $\theta_J$ and construct a prethermal phase diagram in the $(\theta_x, \theta_J)$ parameter space.
To distinguish the rigid subharmonic response characteristic of a DTC from both rapid thermalization and enveloped oscillations emblematic of unstable periodicity, we define the spectral power density $|S_\omega^{A/B}|^2$ at the subharmonic peak $\omega \tau=\pi$ as a prethermal DTC order parameter in either sublattice of the bipartite geometry. 
For the $A$- and $B$-lattices of a decorated hexagonal graph, this gives rise to the diagrams shown in Fig.~\ref{fig:phasediagram}(a-b).

In the absence of interactions ($\theta_J=0$), a nonzero DTC order parameter $|S_{\pi/\tau}|^2$ appears in both lattices only for a perfect periodic drive $\theta_x=\pi$, as the system lacks any mechanism stabilizing the response to an imperfect drive. 
The onset of Ising interactions ($\theta_J<0$) enables self-stabilization of the subharmonic response to increasingly perturbed $\pi$-flips, resulting in a finite range of $\theta_x$ for which the system exhibits stable DTC order (shaded in red). This region, which grows linearly with $|\theta_J|$ in both sublattices, signals the emergence of an interaction-stabilized DTC phase. This behavior is consistent with the presence of a confining potential that inhibits rapid thermalization induced by a weak transverse field~\cite{tindallConfinementTransverseField2024}.
Crucially, stable DTC order at weak interactions does not decay gradually as perturbations $|\theta_x-\pi|$ increase. Instead, it falls off abruptly in a sharp crossover between time-crystalline order and thermalization.

For the decorated hexagonal lattice in Fig.~\ref{fig:phasediagram}(a-b), the stabilizing effect of interactions stagnates around $\theta_J\approx -\pi/3$. Beyond this point, the range of perturbations $|\theta_x-\pi|$ to which the time crystal is robust shrinks until the eventual disappearance of DTC order in the node at $\theta_J =-2\pi/3$, followed by a revival at larger interaction strengths.
This disappearance and recurrence of DTC order as a function of $\theta_J$ is a general feature of decorated lattice structures, as seen from the $B$-sublattice order parameter $|S_{\pi/\tau}^B|^2$ in the thermodynamic limit of a Lieb lattice ($z=4$) and decorated 5-regular graph ($z=5$) shown in Fig.~\ref{fig:phasediagram}(c-d).
In particular, nodes of vanishing DTC order occur at values $\theta_J=-2n\pi/z$, with $n\in \mathbb{N}$ and $z$ the connectivity of the $B$-lattice. This property is readily explained by the fact that the $R_{ZZ}^{jk}(\theta_J)$ interaction terms in Eq.~\eqref{eq:floquet_unitary} effectively rotate each of the involved qubits by an angle $\theta_J$ around the $Z$-axis. Hence, the total rotation of a $z$-connected qubit in the $B$-lattice interacting with its $z$ neighbors in a single Floquet cycle is $z\theta_J$. At $\theta_J =-2n\pi/z$ this reduces to $-2n\pi$, reintroducing the instability of a non-interacting model at $\theta_J=0$.
Following the same reasoning, qubits in the $A$-sublattice of any decorated lattice have two nearest neighbors and thus experience an effective rotation of $2\theta_J$. 
As seen in Fig.~\ref{fig:phasediagram}(a), this results in the disappearance of DTC order in the $A$-sublattice at interaction strengths $\theta_J=-n\pi$.

However, while nodes of the $B$-lattice occur also in the phase diagram of the $A$-lattice and thus destabilize DTC order in the entire system, the $A$-lattice node at $\theta_J = -\pi$ results in a regime of bipartite DTC order: qubits in the $B$-lattice [Fig.~\ref{fig:phasediagram}(b)] exhibit a robust subharmonic response in the limit $\theta_J\to -\pi$, whereas the $A$-lattice [Fig.~\ref{fig:phasediagram}(a)] acquires a quasi-crystalline beating similar to the dynamics in Fig.~\ref{fig:DTC_rigidity}. This is shown by the blue-shaded amplitude of the largest non-subharmonic peak in the spectral density $|S_{\omega}^A|^2$ in Fig.~\ref{fig:phasediagram}(a).
Finally, we note that regimes of bipartite DTC order are not observed in Lieb lattices with even coordination number $z=4$. This follows from the counting argument above, according to which the $B$-lattice exhibits a DTC node at $\theta_J = -n\pi/2$, shown in Fig.~\ref{fig:phasediagram}(c).
The coincidence of the DTC node $\theta_J=-\pi$ with that of the $A$-lattice prevents the emergence of a bipartite DTC phase witnessed in decorated hexagonal and 5-regular graphs.

\, 

\, 

\textbf{\emph{Conclusion.}}~---~
Through tensor network simulations in the thermodynamic limit, we have studied discrete time crystals in periodically driven Ising models on decorated lattices.
We have established the existence of a prethermal phase of time-crystalline order which is exponentially long-lived and decays in a nonperturbative fashion from small deviations from the ideal  $\pi$-periodic drive.
Additionally, we have found that the sensitivity to a longitudinal field, observed on a finite decorated hexagonal lattice in Ref.~\cite{shinjoUnveilingCleanTwodimensional2024}, manifests itself in only one of the two sublattices comprising a generic decorated graph structure. The latter observation implies the existence of a state of bipartite time-crystalline order, in which an extensive part of a bipartite system resisting thermalization displays subharmonic dynamics robust to generic local perturbations. Finally, we have constructed a prethermal DTC phase diagram for the specific cases of a decorated hexagonal lattice, Lieb lattice and decorated 5-regular graph, uncovering a decay and resurrection of both uniform and bipartite prethermal DTC order at interaction strengths determined by the topology of the graph. 

The bipartite discrete time crystals we have introduced bear a resemblance to the notion of boundary time crystals (BTC)~\cite{ieminiBoundaryTimeCrystals2018, piccittoSymmetriesConservedQuantities2021, carolloExactSolutionBoundary2022} and topological Floquet phases of edge states~\cite{zhangDigitalQuantumSimulation2022, bhowmickDiscreteTimeCrystal2023, jiangPrethermalTimeCrystallineCorner2024}.
In contrast to these phenomena, bipartite DTCs as introduced in this work are distinguished by their extensive scaling with system size. 
As such, we anticipate experimental realizations of this phenomenon not only on current digital quantum processors~\cite{kimEvidenceUtilityQuantum2023, aruteQuantumSupremacyUsing2019, boothbyNextGenerationTopologyDWave2020, zhangDigitalQuantumSimulation2022, miTimecrystallineEigenstateOrder2022, kingComputationalSupremacyQuantum2024, xiangLonglivedTopologicalTimecrystalline2024}, but also on analog quantum simulators with periodically driven Ising interactions on non-trivial topologies. Possible candidates for such studies include atomic systems with tunable lattice structures~\cite{taieCoherentDrivingFreezing2015, periwalProgrammableInteractionsEmergent2021, lebratFerrimagnetismUltracoldFermions2024} and condensed matter platforms such as spin impurities in diamond~\cite{choiObservationDiscreteTimecrystalline2017, heExperimentalRealizationDiscrete2024}.
However, further theoretical research is required to elucidate the physical mechanisms underpinning bipartite time-crystallinity, as well as to investigate whether the phenomenology presented here can be extended to the breaking of a continuous time symmetry in driven-dissipative systems~\cite{bucaNonstationaryCoherentQuantum2019, kongkhambutObservationContinuousTime2022,carraro-haddadSolidstateContinuousTime2024,seiboldDissipativeTimeCrystal2020} with a bipartite geometry.

\textit{Acknowledgements.}~---~We thank Flaviano Morone, Michiel Wouters, Kilian Seibold and Berislav Bu\v{c}a for insightful discussions.
L.F. was supported by a Philips postdoctoral fellowship of the Belgian American Educational Foundation. 
J.T. and D.S. are grateful for ongoing support through the Flatiron Institute, a division of the Simons Foundation. D.S. and L.F. are partially supported by AFOSR (Award no. FA9550-21-1-0236) and NSF (Award no. OAC2118310).
The code used to produce the numerical results in this paper was written using the \texttt{ITensorNetworks.jl} package \cite{ITensorNetworksJl2024}, a general purpose and publicly available Julia \cite{bezansonJuliaFreshApproach2017} package for manipulating tensor network states of arbitrary geometry. It is built on top of \texttt{ITensors.jl} \cite{fishmanITensorSoftwareLibrary2022}, which provides the basic tensor operations.
This work was supported in part through the NYU IT High Performance Computing resources, services, and staff expertise.

\bibliography{References}

\clearpage

\input{Supplemental}

\end{document}

%% file: Supplemental.tex
\setcounter{figure}{0}
\setcounter{equation}{0}
\renewcommand{\thefigure}{S\arabic{figure}}
\renewcommand{\theequation}{S\arabic{equation}}


\onecolumngrid

\begin{center}
    \Large{Supplemental material}\\
   \large{to} \\
    \Large \emph{Nonperturbative decay of bipartite discrete time crystals}\\
    \vspace{1em}
    \normalsize{Lennart Fernandes, Joseph Tindall, and Dries Sels}
\end{center}


\section{Convergence of BP-iTNS results}
\label{app:TNS}
\begin{figure}[H]
    \centering
    \includegraphics[width=0.6\linewidth]{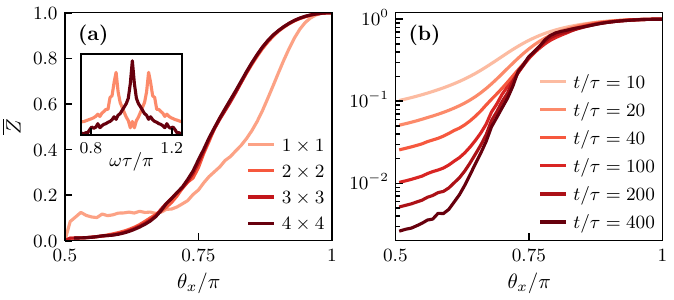}
    \caption{\textbf{Scaling analysis of BP-iTNS results.} (a)~Scaling 
    of the time-averaged magnetization over 100 Floquet cycles with the number of unit cells in a finite decorated hexagonal lattice evaluated at $\chi=80$. The inset shows the same result as Fig.~3(c) evaluated in the bulk of a $111$-site system composed of $4 \times 4$ decorated hexagons. (b)~Convergence of the magnetization in the BP-iTNS with the number of cycles at $\chi=320$.}
    \label{fig:appendix_TNS}
\end{figure}

To validate our BP-iTNS results obtained for the unit cell of an infinite lattice, we provide in Figure~\ref{fig:appendix_TNS}(a) a scaling analysis of the result of Fig.~2(d) in the main text for finite systems composed of $L \times L$ decorated hexagons for $L=\{1,2,3,4\}$, corresponding to $N = \{12, 35, 68, 111\}$ qubits. As proven in more detail in Ref.~[61], this illustrates that the BP approximation converges rapidly with system size in both the time-crystalline and thermalizing regimes. 
The inset shows a repetition of Fig.~3(c) in the main text, but measured in the bulk of a $111$-site system composed of $4 \times 4$ decorated hexagons, illustrating the survival of bipartite DTC order in finite systems.
In infinite lattices represented by the $K_{2,z}$ unit cell, we have confirmed that bipartite DTC order is robust w.r.t. symmetry-broken initial states of the form $\bigotimes_{i=1}^{z+2} |\sigma_i \rangle$, with $\sigma_i = \uparrow\downarrow$ a $Z$-polarized state of a qubit in the unit cell. We note that the employed BP-iTNS approach intrinsically limits our results in the thermodynamic limit to systems with spatial translational invariance.

\,

\begin{figure}[H]
    \centering
    \includegraphics[width=0.6\linewidth]{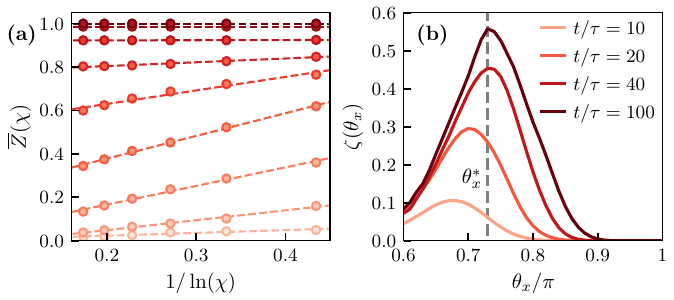}
    \caption{\textbf{Finite entanglement scaling.} (a)~Linear scaling of the time-averaged magnetization during $t/\tau=100$ Floquet cycles with $1/\ln(\chi)$, as in the inset of Fig.~2(d), for $\theta_x/\pi=\{0.6:0.05:1.0\}$ (light to dark). (b)~Time evolution of the extrapolation slope $\zeta(\theta_x)$ as in Fig.~2(e), emphasizing the convergence of the maximum to $\theta_x^*$.}
    \label{fig:appendix_Entanglement}
\end{figure}

As dicussed in the main text, the lifetime of the prethermal DTC scales exponentially with $1/|\theta_x - \pi|$. To illustrate the convergence of our results within a long prethermal stage, we show in Fig.~\ref{fig:appendix_TNS}(b) the mean magnetization in the $N$-qubit system averaged over several numbers $t/\tau$ of Floquet cycles,
\begin{equation}
    \overline{Z} = \frac{\tau}{t N} \sum_{n=1}^{t/\tau} \sum_{j=1}^{N} \langle \psi(n\tau) | (-1)^n Z_{j} | \psi(n\tau) \rangle.
\end{equation}
The thermalizing regime is characterized by a persistent decay of magnetization consistent with the eventual disappearance of order at long times. By contrast, in the DTC regime $\theta_x \to \pi$, $\overline{Z}$ shows no significant decay, highlighting the persistence of DTC order over long time scales despite perturbations.
The field strength $\theta_x^{*}$ of the crossover between these regimes, indicated in Fig.~2(d-e) in the main text, is inferred from the convergence in time of the finite entanglement scaling relation Eq.~(5) in the main text. 
The validity of this linear scaling in $1/\ln(\chi)$, illustrated by the inset of Fig.~2(d) of the main text, is shown in Fig.~\ref{fig:appendix_Entanglement}(a) for a wider range of $\theta_x$ in both the thermalizing and DTC regimes.
As shown in Fig.~\ref{fig:appendix_Entanglement}(b), the extrapolation slope $\zeta(\theta_x)$ reaches a maximum in the intermediary regime, which converges to $\theta_x^{*}\approx 0.73$ with increasing number of Floquet cycles.

\section{State vector simulation}
\label{app:statevector}
\begin{figure}[H]
    \centering
    \includegraphics[width=0.9\linewidth]{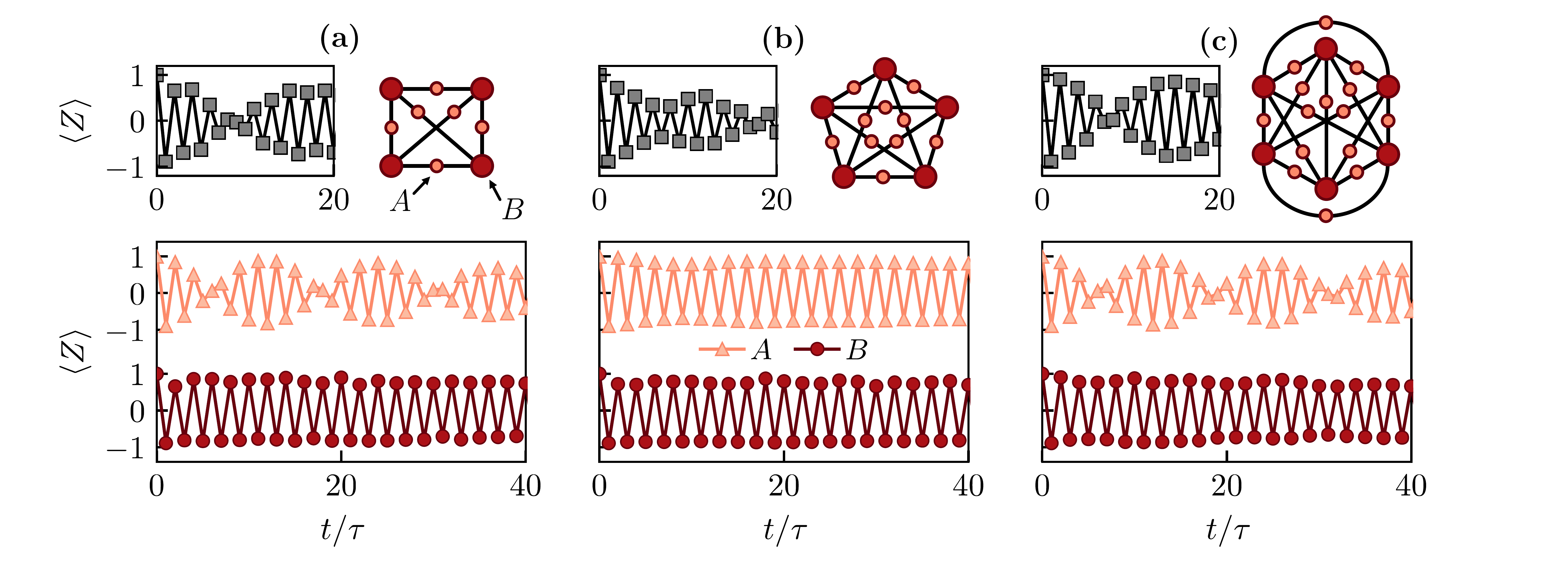}
    \caption{\textbf{Bipartite rigidity in decorated regular graphs.} Exact dynamics of the magnetization $\langle Z \rangle$ generated by the Floquet unitary with rotation angles $(\theta_x, \theta_z) = (0.85\pi, 0.4\pi)$ as in Fig.~3(c), on a $3$-regular (a), $4$-regular (b) and $5$-regular graph. The top row shows the dynamics for the undecorated graphs (i.e., in absence of the orange nodes); the bottom row shows the dynamics on the $A$ and $B$-qubits of the bipartite decorated graphs. The interaction strength $\theta_J$ was taken to be $\pi/2$ (a,c) and $\pi/3$ (b), within the DTC regime of the respective decorated graphs.}
    \label{fig:statevector_simulation}
\end{figure}

For small systems, the dynamics generated by the Floquet unitary can be evaluated exactly by the direct application of one- and two-qubit gates on the state vector expressed in the computational basis of the $2^N$-dimensional Hilbert space. This allows us to verify our conclusions obtained in the BP-iTNS approach with small graphs exhibiting a similar topology.
To validate the existence of bipartite DTCs, we simulate the Floquet dynamics considered in Fig.~3 of the main text on complete $z$-regular graphs of a minimal size for $z=\{3,4,5\}$, with decorated edges. 
Shown in the top row are the magnetization dynamics on any of the equivalent qubits in the undecorated graphs comprising only the red nodes. For Floquet parameters $(\theta_J, \theta_x, \theta_z) = (0.5\pi, 0.85\pi, 0.4\pi)$, the magnetization lacks robust subharmonic order for graphs of any degree, and instead experiences a beating due to the presence of both a transverse and longitudinal field. 
Decorating the edges with additional (orange) qubits gives rise to the bipartite graphs shown in the top row of Fig.~\ref{fig:statevector_simulation}.
For $z=3$ (a) and $z=4$ (b), these graphs have the same degrees of connectivity as the iTNS unit cells shown in Fig.~1(c-d).
As in the main text, the resulting $A$- and $B$-sublattices now exhibit qualitatively different dynamics.
Whereas the dynamics of the magnetization on the decorated qubits is unstable to any longitudinal field perturbation, the subharmonic response on $z$-connected qubits is stabilized for $z=3$ [Fig. \ref{fig:statevector_simulation}(a)] and $z=5$ [Fig. \ref{fig:statevector_simulation}(c)]. By contrast, time-crystalline order in the magnetization of a $z=4$ bipartite graph [Fig. \ref{fig:statevector_simulation}(b)] appears uniformly across both sublattices, as it does in the BP-iTNS approximation of an infinite Lieb lattice.

\pagebreak

\vfill

\vfill

\vfill

\vfill


%% file: main.bbl
\begin{thebibliography}{94}%
\makeatletter
\providecommand \@ifxundefined [1]{%
 \@ifx{#1\undefined}
}%
\providecommand \@ifnum [1]{%
 \ifnum #1\expandafter \@firstoftwo
 \else \expandafter \@secondoftwo
 \fi
}%
\providecommand \@ifx [1]{%
 \ifx #1\expandafter \@firstoftwo
 \else \expandafter \@secondoftwo
 \fi
}%
\providecommand \natexlab [1]{#1}%
\providecommand \enquote  [1]{``#1''}%
\providecommand \bibnamefont  [1]{#1}%
\providecommand \bibfnamefont [1]{#1}%
\providecommand \citenamefont [1]{#1}%
\providecommand \href@noop [0]{\@secondoftwo}%
\providecommand \href [0]{\begingroup \@sanitize@url \@href}%
\providecommand \@href[1]{\@@startlink{#1}\@@href}%
\providecommand \@@href[1]{\endgroup#1\@@endlink}%
\providecommand \@sanitize@url [0]{\catcode `\\12\catcode `\$12\catcode `\&12\catcode `\#12\catcode `\^12\catcode `\_12\catcode `\%12\relax}%
\providecommand \@@startlink[1]{}%
\providecommand \@@endlink[0]{}%
\providecommand \url  [0]{\begingroup\@sanitize@url \@url }%
\providecommand \@url [1]{\endgroup\@href {#1}{\urlprefix }}%
\providecommand \urlprefix  [0]{URL }%
\providecommand \Eprint [0]{\href }%
\providecommand \doibase [0]{https://doi.org/}%
\providecommand \selectlanguage [0]{\@gobble}%
\providecommand \bibinfo  [0]{\@secondoftwo}%
\providecommand \bibfield  [0]{\@secondoftwo}%
\providecommand \translation [1]{[#1]}%
\providecommand \BibitemOpen [0]{}%
\providecommand \bibitemStop [0]{}%
\providecommand \bibitemNoStop [0]{.\EOS\space}%
\providecommand \EOS [0]{\spacefactor3000\relax}%
\providecommand \BibitemShut  [1]{\csname bibitem#1\endcsname}%
\let\auto@bib@innerbib\@empty
\bibitem [{\citenamefont {Wilczek}(2012)}]{wilczekQuantumTimeCrystals2012}%
  \BibitemOpen
  \bibfield  {author} {\bibinfo {author} {\bibfnamefont {F.}~\bibnamefont {Wilczek}},\ }\bibfield  {title} {\bibinfo {title} {Quantum {{Time Crystals}}},\ }\href {https://doi.org/10.1103/PhysRevLett.109.160401} {\bibfield  {journal} {\bibinfo  {journal} {Phys. Rev. Lett.}\ }\textbf {\bibinfo {volume} {109}},\ \bibinfo {pages} {160401} (\bibinfo {year} {2012})}\BibitemShut {NoStop}%
\bibitem [{\citenamefont {Sacha}\ and\ \citenamefont {Zakrzewski}(2018)}]{sachaTimeCrystalsReview2018}%
  \BibitemOpen
  \bibfield  {author} {\bibinfo {author} {\bibfnamefont {K.}~\bibnamefont {Sacha}}\ and\ \bibinfo {author} {\bibfnamefont {J.}~\bibnamefont {Zakrzewski}},\ }\bibfield  {title} {\bibinfo {title} {Time crystals: a review},\ }\href {https://doi.org/10.1088/1361-6633/aa8b38} {\bibfield  {journal} {\bibinfo  {journal} {Rep. Prog. Phys.}\ }\textbf {\bibinfo {volume} {81}},\ \bibinfo {pages} {016401} (\bibinfo {year} {2018})}\BibitemShut {NoStop}%
\bibitem [{\citenamefont {Bruno}(2013)}]{brunoImpossibilitySpontaneouslyRotating2013}%
  \BibitemOpen
  \bibfield  {author} {\bibinfo {author} {\bibfnamefont {P.}~\bibnamefont {Bruno}},\ }\bibfield  {title} {\bibinfo {title} {Impossibility of {{Spontaneously Rotating Time Crystals}}: {{A No-Go Theorem}}},\ }\href {https://doi.org/10.1103/PhysRevLett.111.070402} {\bibfield  {journal} {\bibinfo  {journal} {Phys. Rev. Lett.}\ }\textbf {\bibinfo {volume} {111}},\ \bibinfo {pages} {070402} (\bibinfo {year} {2013})}\BibitemShut {NoStop}%
\bibitem [{\citenamefont {Nozi{\`e}res}(2013)}]{nozieresTimeCrystalsCan2013}%
  \BibitemOpen
  \bibfield  {author} {\bibinfo {author} {\bibfnamefont {P.}~\bibnamefont {Nozi{\`e}res}},\ }\bibfield  {title} {\bibinfo {title} {Time crystals: {{Can}} diamagnetic currents drive a charge density wave into rotation?},\ }\href {https://doi.org/10.1209/0295-5075/103/57008} {\bibfield  {journal} {\bibinfo  {journal} {EPL}\ }\textbf {\bibinfo {volume} {103}},\ \bibinfo {pages} {57008} (\bibinfo {year} {2013})}\BibitemShut {NoStop}%
\bibitem [{\citenamefont {Watanabe}\ and\ \citenamefont {Oshikawa}(2015)}]{watanabeAbsenceQuantumTime2015}%
  \BibitemOpen
  \bibfield  {author} {\bibinfo {author} {\bibfnamefont {H.}~\bibnamefont {Watanabe}}\ and\ \bibinfo {author} {\bibfnamefont {M.}~\bibnamefont {Oshikawa}},\ }\bibfield  {title} {\bibinfo {title} {Absence of {{Quantum Time Crystals}}},\ }\href {https://doi.org/10.1103/PhysRevLett.114.251603} {\bibfield  {journal} {\bibinfo  {journal} {Phys. Rev. Lett.}\ }\textbf {\bibinfo {volume} {114}},\ \bibinfo {pages} {251603} (\bibinfo {year} {2015})}\BibitemShut {NoStop}%
\bibitem [{\citenamefont {Floquet}(1883)}]{floquetEquationsDifferentiellesLineaires1883}%
  \BibitemOpen
  \bibfield  {author} {\bibinfo {author} {\bibfnamefont {G.}~\bibnamefont {Floquet}},\ }\bibfield  {title} {\bibinfo {title} {{Sur les {\'e}quations diff{\'e}rentielles lin{\'e}aires {\`a} coefficients p{\'e}riodiques}},\ }\href {https://doi.org/10.24033/asens.220} {\bibfield  {journal} {\bibinfo  {journal} {Ann. Sci. {\'E}cole Norm. Sup.}\ }\textbf {\bibinfo {volume} {12}},\ \bibinfo {pages} {47} (\bibinfo {year} {1883})}\BibitemShut {NoStop}%
\bibitem [{\citenamefont {Shirley}(1965)}]{shirleySolutionSchrodingerEquation1965}%
  \BibitemOpen
  \bibfield  {author} {\bibinfo {author} {\bibfnamefont {J.~H.}\ \bibnamefont {Shirley}},\ }\bibfield  {title} {\bibinfo {title} {Solution of the {{Schr{\"o}dinger Equation}} with a {{Hamiltonian Periodic}} in {{Time}}},\ }\href {https://doi.org/10.1103/PhysRev.138.B979} {\bibfield  {journal} {\bibinfo  {journal} {Phys. Rev.}\ }\textbf {\bibinfo {volume} {138}},\ \bibinfo {pages} {B979} (\bibinfo {year} {1965})}\BibitemShut {NoStop}%
\bibitem [{\citenamefont {Moessner}\ and\ \citenamefont {Sondhi}(2017)}]{moessnerEquilibrationOrderQuantum2017}%
  \BibitemOpen
  \bibfield  {author} {\bibinfo {author} {\bibfnamefont {R.}~\bibnamefont {Moessner}}\ and\ \bibinfo {author} {\bibfnamefont {S.~L.}\ \bibnamefont {Sondhi}},\ }\bibfield  {title} {\bibinfo {title} {Equilibration and order in quantum {{Floquet}} matter},\ }\href {https://doi.org/10.1038/nphys4106} {\bibfield  {journal} {\bibinfo  {journal} {Nat. Phys.}\ }\textbf {\bibinfo {volume} {13}},\ \bibinfo {pages} {424} (\bibinfo {year} {2017})}\BibitemShut {NoStop}%
\bibitem [{\citenamefont {Sacha}(2015)}]{sachaModelingSpontaneousBreaking2015}%
  \BibitemOpen
  \bibfield  {author} {\bibinfo {author} {\bibfnamefont {K.}~\bibnamefont {Sacha}},\ }\bibfield  {title} {\bibinfo {title} {Modeling spontaneous breaking of time-translation symmetry},\ }\href {https://doi.org/10.1103/PhysRevA.91.033617} {\bibfield  {journal} {\bibinfo  {journal} {Phys. Rev. A}\ }\textbf {\bibinfo {volume} {91}},\ \bibinfo {pages} {033617} (\bibinfo {year} {2015})}\BibitemShut {NoStop}%
\bibitem [{\citenamefont {Else}\ \emph {et~al.}(2020{\natexlab{a}})\citenamefont {Else}, \citenamefont {Monroe}, \citenamefont {Nayak},\ and\ \citenamefont {Yao}}]{elseDiscreteTimeCrystals2020}%
  \BibitemOpen
  \bibfield  {author} {\bibinfo {author} {\bibfnamefont {D.~V.}\ \bibnamefont {Else}}, \bibinfo {author} {\bibfnamefont {C.}~\bibnamefont {Monroe}}, \bibinfo {author} {\bibfnamefont {C.}~\bibnamefont {Nayak}},\ and\ \bibinfo {author} {\bibfnamefont {N.~Y.}\ \bibnamefont {Yao}},\ }\bibfield  {title} {\bibinfo {title} {Discrete {{Time Crystals}}},\ }\href {https://doi.org/10.1146/annurev-conmatphys-031119-050658} {\bibfield  {journal} {\bibinfo  {journal} {Annu. Rev. Condens. Matter Phys.}\ }\textbf {\bibinfo {volume} {11}},\ \bibinfo {pages} {467} (\bibinfo {year} {2020}{\natexlab{a}})}\BibitemShut {NoStop}%
\bibitem [{\citenamefont {Zaletel}\ \emph {et~al.}(2023)\citenamefont {Zaletel}, \citenamefont {Lukin}, \citenamefont {Monroe}, \citenamefont {Nayak}, \citenamefont {Wilczek},\ and\ \citenamefont {Yao}}]{zaletelColloquiumQuantumClassical2023}%
  \BibitemOpen
  \bibfield  {author} {\bibinfo {author} {\bibfnamefont {M.~P.}\ \bibnamefont {Zaletel}}, \bibinfo {author} {\bibfnamefont {M.}~\bibnamefont {Lukin}}, \bibinfo {author} {\bibfnamefont {C.}~\bibnamefont {Monroe}}, \bibinfo {author} {\bibfnamefont {C.}~\bibnamefont {Nayak}}, \bibinfo {author} {\bibfnamefont {F.}~\bibnamefont {Wilczek}},\ and\ \bibinfo {author} {\bibfnamefont {N.~Y.}\ \bibnamefont {Yao}},\ }\bibfield  {title} {\bibinfo {title} {{\emph{Colloquium}} : {{Quantum}} and classical discrete time crystals},\ }\href {https://doi.org/10.1103/RevModPhys.95.031001} {\bibfield  {journal} {\bibinfo  {journal} {Rev. Mod. Phys.}\ }\textbf {\bibinfo {volume} {95}},\ \bibinfo {pages} {031001} (\bibinfo {year} {2023})}\BibitemShut {NoStop}%
\bibitem [{\citenamefont {Turner}\ \emph {et~al.}(2018)\citenamefont {Turner}, \citenamefont {Michailidis}, \citenamefont {Abanin}, \citenamefont {Serbyn},\ and\ \citenamefont {Papi{\'c}}}]{turnerWeakErgodicityBreaking2018}%
  \BibitemOpen
  \bibfield  {author} {\bibinfo {author} {\bibfnamefont {C.~J.}\ \bibnamefont {Turner}}, \bibinfo {author} {\bibfnamefont {A.~A.}\ \bibnamefont {Michailidis}}, \bibinfo {author} {\bibfnamefont {D.~A.}\ \bibnamefont {Abanin}}, \bibinfo {author} {\bibfnamefont {M.}~\bibnamefont {Serbyn}},\ and\ \bibinfo {author} {\bibfnamefont {Z.}~\bibnamefont {Papi{\'c}}},\ }\bibfield  {title} {\bibinfo {title} {Weak ergodicity breaking from quantum many-body scars},\ }\href {https://doi.org/10.1038/s41567-018-0137-5} {\bibfield  {journal} {\bibinfo  {journal} {Nat. Phys.}\ }\textbf {\bibinfo {volume} {14}},\ \bibinfo {pages} {745} (\bibinfo {year} {2018})}\BibitemShut {NoStop}%
\bibitem [{\citenamefont {Bluvstein}\ \emph {et~al.}(2021)\citenamefont {Bluvstein} \emph {et~al.}}]{bluvsteinControllingQuantumManybody2021}%
  \BibitemOpen
  \bibfield  {author} {\bibinfo {author} {\bibfnamefont {D.}~\bibnamefont {Bluvstein}} \emph {et~al.},\ }\bibfield  {title} {\bibinfo {title} {Controlling quantum many-body dynamics in driven {{Rydberg}} atom arrays},\ }\href {https://doi.org/10.1126/science.abg2530} {\bibfield  {journal} {\bibinfo  {journal} {Science}\ }\textbf {\bibinfo {volume} {371}},\ \bibinfo {pages} {1355} (\bibinfo {year} {2021})}\BibitemShut {NoStop}%
\bibitem [{\citenamefont {Maskara}\ \emph {et~al.}(2021)\citenamefont {Maskara}, \citenamefont {Michailidis}, \citenamefont {Ho}, \citenamefont {Bluvstein}, \citenamefont {Choi}, \citenamefont {Lukin},\ and\ \citenamefont {Serbyn}}]{maskaraDiscreteTimeCrystallineOrder2021}%
  \BibitemOpen
  \bibfield  {author} {\bibinfo {author} {\bibfnamefont {N.}~\bibnamefont {Maskara}}, \bibinfo {author} {\bibfnamefont {A.~A.}\ \bibnamefont {Michailidis}}, \bibinfo {author} {\bibfnamefont {W.~W.}\ \bibnamefont {Ho}}, \bibinfo {author} {\bibfnamefont {D.}~\bibnamefont {Bluvstein}}, \bibinfo {author} {\bibfnamefont {S.}~\bibnamefont {Choi}}, \bibinfo {author} {\bibfnamefont {M.~D.}\ \bibnamefont {Lukin}},\ and\ \bibinfo {author} {\bibfnamefont {M.}~\bibnamefont {Serbyn}},\ }\bibfield  {title} {\bibinfo {title} {Discrete {{Time-Crystalline Order Enabled}} by {{Quantum Many-Body Scars}}: {{Entanglement Steering}} via {{Periodic Driving}}},\ }\href {https://doi.org/10.1103/PhysRevLett.127.090602} {\bibfield  {journal} {\bibinfo  {journal} {Phys. Rev. Lett.}\ }\textbf {\bibinfo {volume} {127}},\ \bibinfo {pages} {090602} (\bibinfo {year} {2021})}\BibitemShut {NoStop}%
\bibitem [{\citenamefont {Huang}\ \emph {et~al.}(2022)\citenamefont {Huang}, \citenamefont {Leung}, \citenamefont {{Stamper-Kurn}},\ and\ \citenamefont {Liu}}]{huangDiscreteTimeCrystals2022}%
  \BibitemOpen
  \bibfield  {author} {\bibinfo {author} {\bibfnamefont {B.}~\bibnamefont {Huang}}, \bibinfo {author} {\bibfnamefont {T.-H.}\ \bibnamefont {Leung}}, \bibinfo {author} {\bibfnamefont {D.~M.}\ \bibnamefont {{Stamper-Kurn}}},\ and\ \bibinfo {author} {\bibfnamefont {W.~V.}\ \bibnamefont {Liu}},\ }\bibfield  {title} {\bibinfo {title} {Discrete {{Time Crystals Enforced}} by {{Floquet-Bloch Scars}}},\ }\href {https://doi.org/10.1103/PhysRevLett.129.133001} {\bibfield  {journal} {\bibinfo  {journal} {Phys. Rev. Lett.}\ }\textbf {\bibinfo {volume} {129}},\ \bibinfo {pages} {133001} (\bibinfo {year} {2022})}\BibitemShut {NoStop}%
\bibitem [{\citenamefont {Nandkishore}\ and\ \citenamefont {Huse}(2015)}]{nandkishoreManyBodyLocalizationThermalization2015}%
  \BibitemOpen
  \bibfield  {author} {\bibinfo {author} {\bibfnamefont {R.}~\bibnamefont {Nandkishore}}\ and\ \bibinfo {author} {\bibfnamefont {D.~A.}\ \bibnamefont {Huse}},\ }\bibfield  {title} {\bibinfo {title} {Many-{{Body Localization}} and {{Thermalization}} in {{Quantum Statistical Mechanics}}},\ }\href {https://doi.org/10.1146/annurev-conmatphys-031214-014726} {\bibfield  {journal} {\bibinfo  {journal} {Annu. Rev. Condens. Matter Phys.}\ }\textbf {\bibinfo {volume} {6}},\ \bibinfo {pages} {15} (\bibinfo {year} {2015})}\BibitemShut {NoStop}%
\bibitem [{\citenamefont {Abanin}\ \emph {et~al.}(2019)\citenamefont {Abanin}, \citenamefont {Altman}, \citenamefont {Bloch},\ and\ \citenamefont {Serbyn}}]{abaninColloquiumManybodyLocalization2019}%
  \BibitemOpen
  \bibfield  {author} {\bibinfo {author} {\bibfnamefont {D.~A.}\ \bibnamefont {Abanin}}, \bibinfo {author} {\bibfnamefont {E.}~\bibnamefont {Altman}}, \bibinfo {author} {\bibfnamefont {I.}~\bibnamefont {Bloch}},\ and\ \bibinfo {author} {\bibfnamefont {M.}~\bibnamefont {Serbyn}},\ }\bibfield  {title} {\bibinfo {title} {{\emph{Colloquium}}: {{Many-body}} localization, thermalization, and entanglement},\ }\href {https://doi.org/10.1103/RevModPhys.91.021001} {\bibfield  {journal} {\bibinfo  {journal} {Rev. Mod. Phys.}\ }\textbf {\bibinfo {volume} {91}},\ \bibinfo {pages} {021001} (\bibinfo {year} {2019})}\BibitemShut {NoStop}%
\bibitem [{\citenamefont {Else}\ \emph {et~al.}(2016)\citenamefont {Else}, \citenamefont {Bauer},\ and\ \citenamefont {Nayak}}]{elseFloquetTimeCrystals2016}%
  \BibitemOpen
  \bibfield  {author} {\bibinfo {author} {\bibfnamefont {D.~V.}\ \bibnamefont {Else}}, \bibinfo {author} {\bibfnamefont {B.}~\bibnamefont {Bauer}},\ and\ \bibinfo {author} {\bibfnamefont {C.}~\bibnamefont {Nayak}},\ }\bibfield  {title} {\bibinfo {title} {Floquet {{Time Crystals}}},\ }\href {https://doi.org/10.1103/PhysRevLett.117.090402} {\bibfield  {journal} {\bibinfo  {journal} {Phys. Rev. Lett.}\ }\textbf {\bibinfo {volume} {117}},\ \bibinfo {pages} {090402} (\bibinfo {year} {2016})}\BibitemShut {NoStop}%
\bibitem [{\citenamefont {Khemani}\ \emph {et~al.}(2016)\citenamefont {Khemani}, \citenamefont {Lazarides}, \citenamefont {Moessner},\ and\ \citenamefont {Sondhi}}]{khemaniPhaseStructureDriven2016}%
  \BibitemOpen
  \bibfield  {author} {\bibinfo {author} {\bibfnamefont {V.}~\bibnamefont {Khemani}}, \bibinfo {author} {\bibfnamefont {A.}~\bibnamefont {Lazarides}}, \bibinfo {author} {\bibfnamefont {R.}~\bibnamefont {Moessner}},\ and\ \bibinfo {author} {\bibfnamefont {S.~L.}\ \bibnamefont {Sondhi}},\ }\bibfield  {title} {\bibinfo {title} {Phase {{Structure}} of {{Driven Quantum Systems}}},\ }\href {https://doi.org/10.1103/PhysRevLett.116.250401} {\bibfield  {journal} {\bibinfo  {journal} {Phys. Rev. Lett.}\ }\textbf {\bibinfo {volume} {116}},\ \bibinfo {pages} {250401} (\bibinfo {year} {2016})}\BibitemShut {NoStop}%
\bibitem [{\citenamefont {Yao}\ \emph {et~al.}(2017)\citenamefont {Yao}, \citenamefont {Potter}, \citenamefont {Potirniche},\ and\ \citenamefont {Vishwanath}}]{yaoDiscreteTimeCrystals2017}%
  \BibitemOpen
  \bibfield  {author} {\bibinfo {author} {\bibfnamefont {N.~Y.}\ \bibnamefont {Yao}}, \bibinfo {author} {\bibfnamefont {A.~C.}\ \bibnamefont {Potter}}, \bibinfo {author} {\bibfnamefont {I.-D.}\ \bibnamefont {Potirniche}},\ and\ \bibinfo {author} {\bibfnamefont {A.}~\bibnamefont {Vishwanath}},\ }\bibfield  {title} {\bibinfo {title} {Discrete {{Time Crystals}}: {{Rigidity}}, {{Criticality}}, and {{Realizations}}},\ }\href {https://doi.org/10.1103/PhysRevLett.118.030401} {\bibfield  {journal} {\bibinfo  {journal} {Phys. Rev. Lett.}\ }\textbf {\bibinfo {volume} {118}},\ \bibinfo {pages} {030401} (\bibinfo {year} {2017})}\BibitemShut {NoStop}%
\bibitem [{\citenamefont {Zhang}\ \emph {et~al.}(2017)\citenamefont {Zhang} \emph {et~al.}}]{zhangObservationDiscreteTime2017}%
  \BibitemOpen
  \bibfield  {author} {\bibinfo {author} {\bibfnamefont {J.}~\bibnamefont {Zhang}} \emph {et~al.},\ }\bibfield  {title} {\bibinfo {title} {Observation of a discrete time crystal},\ }\href {https://doi.org/10.1038/nature21413} {\bibfield  {journal} {\bibinfo  {journal} {Nature}\ }\textbf {\bibinfo {volume} {543}},\ \bibinfo {pages} {217} (\bibinfo {year} {2017})}\BibitemShut {NoStop}%
\bibitem [{\citenamefont {Choi}\ \emph {et~al.}(2017)\citenamefont {Choi} \emph {et~al.}}]{choiObservationDiscreteTimecrystalline2017}%
  \BibitemOpen
  \bibfield  {author} {\bibinfo {author} {\bibfnamefont {S.}~\bibnamefont {Choi}} \emph {et~al.},\ }\bibfield  {title} {\bibinfo {title} {Observation of discrete time-crystalline order in a disordered dipolar many-body system},\ }\href {https://doi.org/10.1038/nature21426} {\bibfield  {journal} {\bibinfo  {journal} {Nature}\ }\textbf {\bibinfo {volume} {543}},\ \bibinfo {pages} {221} (\bibinfo {year} {2017})}\BibitemShut {NoStop}%
\bibitem [{\citenamefont {Frey}\ and\ \citenamefont {Rachel}(2022)}]{freyRealizationDiscreteTime2022}%
  \BibitemOpen
  \bibfield  {author} {\bibinfo {author} {\bibfnamefont {P.}~\bibnamefont {Frey}}\ and\ \bibinfo {author} {\bibfnamefont {S.}~\bibnamefont {Rachel}},\ }\bibfield  {title} {\bibinfo {title} {Realization of a discrete time crystal on 57 qubits of a quantum computer},\ }\href {https://doi.org/10.1126/sciadv.abm7652} {\bibfield  {journal} {\bibinfo  {journal} {Sci. Adv.}\ }\textbf {\bibinfo {volume} {8}},\ \bibinfo {pages} {eabm7652} (\bibinfo {year} {2022})}\BibitemShut {NoStop}%
\bibitem [{\citenamefont {Schulz}\ \emph {et~al.}(2019)\citenamefont {Schulz}, \citenamefont {Hooley}, \citenamefont {Moessner},\ and\ \citenamefont {Pollmann}}]{schulzStarkManyBodyLocalization2019}%
  \BibitemOpen
  \bibfield  {author} {\bibinfo {author} {\bibfnamefont {M.}~\bibnamefont {Schulz}}, \bibinfo {author} {\bibfnamefont {C.}~\bibnamefont {Hooley}}, \bibinfo {author} {\bibfnamefont {R.}~\bibnamefont {Moessner}},\ and\ \bibinfo {author} {\bibfnamefont {F.}~\bibnamefont {Pollmann}},\ }\bibfield  {title} {\bibinfo {title} {Stark {Many}-{Body} {Localization}},\ }\href {https://doi.org/10.1103/PhysRevLett.122.040606} {\bibfield  {journal} {\bibinfo  {journal} {Phys. Rev. Lett.}\ }\textbf {\bibinfo {volume} {122}},\ \bibinfo {pages} {040606} (\bibinfo {year} {2019})}\BibitemShut {NoStop}%
\bibitem [{\citenamefont {Van~Nieuwenburg}\ \emph {et~al.}(2019)\citenamefont {Van~Nieuwenburg}, \citenamefont {Baum},\ and\ \citenamefont {Refael}}]{vannieuwenburgBlochOscillationsManybody2019}%
  \BibitemOpen
  \bibfield  {author} {\bibinfo {author} {\bibfnamefont {E.}~\bibnamefont {Van~Nieuwenburg}}, \bibinfo {author} {\bibfnamefont {Y.}~\bibnamefont {Baum}},\ and\ \bibinfo {author} {\bibfnamefont {G.}~\bibnamefont {Refael}},\ }\bibfield  {title} {\bibinfo {title} {From {Bloch} oscillations to many-body localization in clean interacting systems},\ }\href {https://doi.org/10.1073/pnas.1819316116} {\bibfield  {journal} {\bibinfo  {journal} {PNAS}\ }\textbf {\bibinfo {volume} {116}},\ \bibinfo {pages} {9269} (\bibinfo {year} {2019})}\BibitemShut {NoStop}%
\bibitem [{\citenamefont {Kshetrimayum}\ \emph {et~al.}(2020)\citenamefont {Kshetrimayum}, \citenamefont {Eisert},\ and\ \citenamefont {Kennes}}]{kshetrimayumStarkTimeCrystals2020}%
  \BibitemOpen
  \bibfield  {author} {\bibinfo {author} {\bibfnamefont {A.}~\bibnamefont {Kshetrimayum}}, \bibinfo {author} {\bibfnamefont {J.}~\bibnamefont {Eisert}},\ and\ \bibinfo {author} {\bibfnamefont {D.~M.}\ \bibnamefont {Kennes}},\ }\bibfield  {title} {\bibinfo {title} {Stark time crystals: {Symmetry} breaking in space and time},\ }\href {https://doi.org/10.1103/PhysRevB.102.195116} {\bibfield  {journal} {\bibinfo  {journal} {Phys. Rev. B}\ }\textbf {\bibinfo {volume} {102}},\ \bibinfo {pages} {195116} (\bibinfo {year} {2020})}\BibitemShut {NoStop}%
\bibitem [{\citenamefont {Liu}\ \emph {et~al.}(2023)\citenamefont {Liu}, \citenamefont {Zhang}, \citenamefont {Hsieh}, \citenamefont {Zhang},\ and\ \citenamefont {Yao}}]{liuDiscreteTimeCrystal2023}%
  \BibitemOpen
  \bibfield  {author} {\bibinfo {author} {\bibfnamefont {S.}~\bibnamefont {Liu}}, \bibinfo {author} {\bibfnamefont {S.-X.}\ \bibnamefont {Zhang}}, \bibinfo {author} {\bibfnamefont {C.-Y.}\ \bibnamefont {Hsieh}}, \bibinfo {author} {\bibfnamefont {S.}~\bibnamefont {Zhang}},\ and\ \bibinfo {author} {\bibfnamefont {H.}~\bibnamefont {Yao}},\ }\bibfield  {title} {\bibinfo {title} {Discrete {Time} {Crystal} {Enabled} by {Stark} {Many}-{Body} {Localization}},\ }\href {https://doi.org/10.1103/PhysRevLett.130.120403} {\bibfield  {journal} {\bibinfo  {journal} {Phys. Rev. Lett.}\ }\textbf {\bibinfo {volume} {130}},\ \bibinfo {pages} {120403} (\bibinfo {year} {2023})}\BibitemShut {NoStop}%
\bibitem [{\citenamefont {Smith}\ \emph {et~al.}(2017)\citenamefont {Smith}, \citenamefont {Knolle}, \citenamefont {Kovrizhin},\ and\ \citenamefont {Moessner}}]{smithDisorderFreeLocalization2017}%
  \BibitemOpen
  \bibfield  {author} {\bibinfo {author} {\bibfnamefont {A.}~\bibnamefont {Smith}}, \bibinfo {author} {\bibfnamefont {J.}~\bibnamefont {Knolle}}, \bibinfo {author} {\bibfnamefont {D.}~\bibnamefont {Kovrizhin}},\ and\ \bibinfo {author} {\bibfnamefont {R.}~\bibnamefont {Moessner}},\ }\bibfield  {title} {\bibinfo {title} {Disorder-{Free} {Localization}},\ }\href {https://doi.org/10.1103/PhysRevLett.118.266601} {\bibfield  {journal} {\bibinfo  {journal} {Phys. Rev. Lett.}\ }\textbf {\bibinfo {volume} {118}},\ \bibinfo {pages} {266601} (\bibinfo {year} {2017})}\BibitemShut {NoStop}%
\bibitem [{\citenamefont {Brenes}\ \emph {et~al.}(2018)\citenamefont {Brenes}, \citenamefont {Dalmonte}, \citenamefont {Heyl},\ and\ \citenamefont {Scardicchio}}]{brenesManyBodyLocalizationDynamics2018}%
  \BibitemOpen
  \bibfield  {author} {\bibinfo {author} {\bibfnamefont {M.}~\bibnamefont {Brenes}}, \bibinfo {author} {\bibfnamefont {M.}~\bibnamefont {Dalmonte}}, \bibinfo {author} {\bibfnamefont {M.}~\bibnamefont {Heyl}},\ and\ \bibinfo {author} {\bibfnamefont {A.}~\bibnamefont {Scardicchio}},\ }\bibfield  {title} {\bibinfo {title} {Many-{Body} {Localization} {Dynamics} from {Gauge} {Invariance}},\ }\href {https://doi.org/10.1103/PhysRevLett.120.030601} {\bibfield  {journal} {\bibinfo  {journal} {Phys. Rev. Lett.}\ }\textbf {\bibinfo {volume} {120}},\ \bibinfo {pages} {030601} (\bibinfo {year} {2018})}\BibitemShut {NoStop}%
\bibitem [{\citenamefont {Russomanno}\ \emph {et~al.}(2020)\citenamefont {Russomanno}, \citenamefont {Notarnicola}, \citenamefont {Surace}, \citenamefont {Fazio}, \citenamefont {Dalmonte},\ and\ \citenamefont {Heyl}}]{russomannoHomogeneousFloquetTime2020}%
  \BibitemOpen
  \bibfield  {author} {\bibinfo {author} {\bibfnamefont {A.}~\bibnamefont {Russomanno}}, \bibinfo {author} {\bibfnamefont {S.}~\bibnamefont {Notarnicola}}, \bibinfo {author} {\bibfnamefont {F.~M.}\ \bibnamefont {Surace}}, \bibinfo {author} {\bibfnamefont {R.}~\bibnamefont {Fazio}}, \bibinfo {author} {\bibfnamefont {M.}~\bibnamefont {Dalmonte}},\ and\ \bibinfo {author} {\bibfnamefont {M.}~\bibnamefont {Heyl}},\ }\bibfield  {title} {\bibinfo {title} {Homogeneous {Floquet} time crystal protected by gauge invariance},\ }\href {https://doi.org/10.1103/PhysRevResearch.2.012003} {\bibfield  {journal} {\bibinfo  {journal} {Phys. Rev. Res.}\ }\textbf {\bibinfo {volume} {2}},\ \bibinfo {pages} {012003} (\bibinfo {year} {2020})}\BibitemShut {NoStop}%
\bibitem [{\citenamefont {Bukov}\ \emph {et~al.}(2015)\citenamefont {Bukov}, \citenamefont {Gopalakrishnan}, \citenamefont {Knap},\ and\ \citenamefont {Demler}}]{bukovPrethermalFloquetSteady2015}%
  \BibitemOpen
  \bibfield  {author} {\bibinfo {author} {\bibfnamefont {M.}~\bibnamefont {Bukov}}, \bibinfo {author} {\bibfnamefont {S.}~\bibnamefont {Gopalakrishnan}}, \bibinfo {author} {\bibfnamefont {M.}~\bibnamefont {Knap}},\ and\ \bibinfo {author} {\bibfnamefont {E.}~\bibnamefont {Demler}},\ }\bibfield  {title} {\bibinfo {title} {Prethermal {Floquet} {Steady} {States} and {Instabilities} in the {Periodically} {Driven}, {Weakly} {Interacting} {Bose}-{Hubbard} {Model}},\ }\href {https://doi.org/10.1103/PhysRevLett.115.205301} {\bibfield  {journal} {\bibinfo  {journal} {Phys. Rev. Lett.}\ }\textbf {\bibinfo {volume} {115}},\ \bibinfo {pages} {205301} (\bibinfo {year} {2015})}\BibitemShut {NoStop}%
\bibitem [{\citenamefont {Weidinger}\ and\ \citenamefont {Knap}(2017)}]{weidingerFloquetPrethermalizationRegimes2017}%
  \BibitemOpen
  \bibfield  {author} {\bibinfo {author} {\bibfnamefont {S.~A.}\ \bibnamefont {Weidinger}}\ and\ \bibinfo {author} {\bibfnamefont {M.}~\bibnamefont {Knap}},\ }\bibfield  {title} {\bibinfo {title} {Floquet prethermalization and regimes of heating in a periodically driven, interacting quantum system},\ }\href {https://doi.org/10.1038/srep45382} {\bibfield  {journal} {\bibinfo  {journal} {Sci. Rep.}\ }\textbf {\bibinfo {volume} {7}},\ \bibinfo {pages} {45382} (\bibinfo {year} {2017})}\BibitemShut {NoStop}%
\bibitem [{\citenamefont {Rubio-Abadal}\ \emph {et~al.}(2020)\citenamefont {Rubio-Abadal}, \citenamefont {Ippoliti}, \citenamefont {Hollerith}, \citenamefont {Wei}, \citenamefont {Rui}, \citenamefont {Sondhi}, \citenamefont {Khemani}, \citenamefont {Gross},\ and\ \citenamefont {Bloch}}]{rubio-abadalFloquetPrethermalizationBoseHubbard2020}%
  \BibitemOpen
  \bibfield  {author} {\bibinfo {author} {\bibfnamefont {A.}~\bibnamefont {Rubio-Abadal}}, \bibinfo {author} {\bibfnamefont {M.}~\bibnamefont {Ippoliti}}, \bibinfo {author} {\bibfnamefont {S.}~\bibnamefont {Hollerith}}, \bibinfo {author} {\bibfnamefont {D.}~\bibnamefont {Wei}}, \bibinfo {author} {\bibfnamefont {J.}~\bibnamefont {Rui}}, \bibinfo {author} {\bibfnamefont {S.}~\bibnamefont {Sondhi}}, \bibinfo {author} {\bibfnamefont {V.}~\bibnamefont {Khemani}}, \bibinfo {author} {\bibfnamefont {C.}~\bibnamefont {Gross}},\ and\ \bibinfo {author} {\bibfnamefont {I.}~\bibnamefont {Bloch}},\ }\bibfield  {title} {\bibinfo {title} {Floquet {Prethermalization} in a {Bose}-{Hubbard} {System}},\ }\href {https://doi.org/10.1103/PhysRevX.10.021044} {\bibfield  {journal} {\bibinfo  {journal} {Phys. Rev. X}\ }\textbf {\bibinfo {volume} {10}},\ \bibinfo {pages} {021044} (\bibinfo {year} {2020})}\BibitemShut {NoStop}%
\bibitem [{\citenamefont {Fleckenstein}\ and\ \citenamefont {Bukov}(2021)}]{fleckensteinThermalizationPrethermalizationPeriodically2021}%
  \BibitemOpen
  \bibfield  {author} {\bibinfo {author} {\bibfnamefont {C.}~\bibnamefont {Fleckenstein}}\ and\ \bibinfo {author} {\bibfnamefont {M.}~\bibnamefont {Bukov}},\ }\bibfield  {title} {\bibinfo {title} {Thermalization and prethermalization in periodically kicked quantum spin chains},\ }\href {https://doi.org/10.1103/PhysRevB.103.144307} {\bibfield  {journal} {\bibinfo  {journal} {Phys. Rev. B}\ }\textbf {\bibinfo {volume} {103}},\ \bibinfo {pages} {144307} (\bibinfo {year} {2021})}\BibitemShut {NoStop}%
\bibitem [{\citenamefont {Ho}\ \emph {et~al.}(2023)\citenamefont {Ho}, \citenamefont {Mori}, \citenamefont {Abanin},\ and\ \citenamefont {Dalla~Torre}}]{hoQuantumClassicalFloquet2023}%
  \BibitemOpen
  \bibfield  {author} {\bibinfo {author} {\bibfnamefont {W.~W.}\ \bibnamefont {Ho}}, \bibinfo {author} {\bibfnamefont {T.}~\bibnamefont {Mori}}, \bibinfo {author} {\bibfnamefont {D.~A.}\ \bibnamefont {Abanin}},\ and\ \bibinfo {author} {\bibfnamefont {E.~G.}\ \bibnamefont {Dalla~Torre}},\ }\bibfield  {title} {\bibinfo {title} {Quantum and classical {Floquet} prethermalization},\ }\href {https://doi.org/10.1016/j.aop.2023.169297} {\bibfield  {journal} {\bibinfo  {journal} {Ann. Phys.}\ }\textbf {\bibinfo {volume} {454}},\ \bibinfo {pages} {169297} (\bibinfo {year} {2023})}\BibitemShut {NoStop}%
\bibitem [{\citenamefont {Abanin}\ \emph {et~al.}(2015)\citenamefont {Abanin}, \citenamefont {De~Roeck},\ and\ \citenamefont {Huveneers}}]{abaninExponentiallySlowHeating2015}%
  \BibitemOpen
  \bibfield  {author} {\bibinfo {author} {\bibfnamefont {D.~A.}\ \bibnamefont {Abanin}}, \bibinfo {author} {\bibfnamefont {W.}~\bibnamefont {De~Roeck}},\ and\ \bibinfo {author} {\bibfnamefont {F.}~\bibnamefont {Huveneers}},\ }\bibfield  {title} {\bibinfo {title} {Exponentially {{Slow Heating}} in {{Periodically Driven Many-Body Systems}}},\ }\href {https://doi.org/10.1103/PhysRevLett.115.256803} {\bibfield  {journal} {\bibinfo  {journal} {Phys. Rev. Lett.}\ }\textbf {\bibinfo {volume} {115}},\ \bibinfo {pages} {256803} (\bibinfo {year} {2015})}\BibitemShut {NoStop}%
\bibitem [{\citenamefont {Mori}\ \emph {et~al.}(2016)\citenamefont {Mori}, \citenamefont {Kuwahara},\ and\ \citenamefont {Saito}}]{moriRigorousBoundEnergy2016}%
  \BibitemOpen
  \bibfield  {author} {\bibinfo {author} {\bibfnamefont {T.}~\bibnamefont {Mori}}, \bibinfo {author} {\bibfnamefont {T.}~\bibnamefont {Kuwahara}},\ and\ \bibinfo {author} {\bibfnamefont {K.}~\bibnamefont {Saito}},\ }\bibfield  {title} {\bibinfo {title} {Rigorous {{Bound}} on {{Energy Absorption}} and {{Generic Relaxation}} in {{Periodically Driven Quantum Systems}}},\ }\href {https://doi.org/10.1103/PhysRevLett.116.120401} {\bibfield  {journal} {\bibinfo  {journal} {Phys. Rev. Lett.}\ }\textbf {\bibinfo {volume} {116}},\ \bibinfo {pages} {120401} (\bibinfo {year} {2016})}\BibitemShut {NoStop}%
\bibitem [{\citenamefont {Abanin}\ \emph {et~al.}(2017)\citenamefont {Abanin}, \citenamefont {De~Roeck}, \citenamefont {Ho},\ and\ \citenamefont {Huveneers}}]{abaninEffectiveHamiltoniansPrethermalization2017}%
  \BibitemOpen
  \bibfield  {author} {\bibinfo {author} {\bibfnamefont {D.~A.}\ \bibnamefont {Abanin}}, \bibinfo {author} {\bibfnamefont {W.}~\bibnamefont {De~Roeck}}, \bibinfo {author} {\bibfnamefont {W.~W.}\ \bibnamefont {Ho}},\ and\ \bibinfo {author} {\bibfnamefont {F.}~\bibnamefont {Huveneers}},\ }\bibfield  {title} {\bibinfo {title} {Effective {{Hamiltonians}}, prethermalization, and slow energy absorption in periodically driven many-body systems},\ }\href {https://doi.org/10.1103/PhysRevB.95.014112} {\bibfield  {journal} {\bibinfo  {journal} {Phys. Rev. B}\ }\textbf {\bibinfo {volume} {95}},\ \bibinfo {pages} {014112} (\bibinfo {year} {2017})}\BibitemShut {NoStop}%
\bibitem [{\citenamefont {Else}\ \emph {et~al.}(2017)\citenamefont {Else}, \citenamefont {Bauer},\ and\ \citenamefont {Nayak}}]{elsePrethermalPhasesMatter2017}%
  \BibitemOpen
  \bibfield  {author} {\bibinfo {author} {\bibfnamefont {D.~V.}\ \bibnamefont {Else}}, \bibinfo {author} {\bibfnamefont {B.}~\bibnamefont {Bauer}},\ and\ \bibinfo {author} {\bibfnamefont {C.}~\bibnamefont {Nayak}},\ }\bibfield  {title} {\bibinfo {title} {Prethermal {{Phases}} of {{Matter Protected}} by {{Time-Translation Symmetry}}},\ }\href {https://doi.org/10.1103/PhysRevX.7.011026} {\bibfield  {journal} {\bibinfo  {journal} {Phys. Rev. X}\ }\textbf {\bibinfo {volume} {7}},\ \bibinfo {pages} {011026} (\bibinfo {year} {2017})}\BibitemShut {NoStop}%
\bibitem [{\citenamefont {Zeng}\ and\ \citenamefont {Sheng}(2017)}]{zengPrethermalTimeCrystals2017}%
  \BibitemOpen
  \bibfield  {author} {\bibinfo {author} {\bibfnamefont {T.-S.}\ \bibnamefont {Zeng}}\ and\ \bibinfo {author} {\bibfnamefont {D.~N.}\ \bibnamefont {Sheng}},\ }\bibfield  {title} {\bibinfo {title} {Prethermal time crystals in a one-dimensional periodically driven {Floquet} system},\ }\href {https://doi.org/10.1103/PhysRevB.96.094202} {\bibfield  {journal} {\bibinfo  {journal} {Phys. Rev. B}\ }\textbf {\bibinfo {volume} {96}},\ \bibinfo {pages} {094202} (\bibinfo {year} {2017})}\BibitemShut {NoStop}%
\bibitem [{\citenamefont {Luitz}\ \emph {et~al.}(2020)\citenamefont {Luitz}, \citenamefont {Moessner}, \citenamefont {Sondhi},\ and\ \citenamefont {Khemani}}]{luitzPrethermalizationTemperature2020}%
  \BibitemOpen
  \bibfield  {author} {\bibinfo {author} {\bibfnamefont {D.~J.}\ \bibnamefont {Luitz}}, \bibinfo {author} {\bibfnamefont {R.}~\bibnamefont {Moessner}}, \bibinfo {author} {\bibfnamefont {S.}~\bibnamefont {Sondhi}},\ and\ \bibinfo {author} {\bibfnamefont {V.}~\bibnamefont {Khemani}},\ }\bibfield  {title} {\bibinfo {title} {Prethermalization without {Temperature}},\ }\href {https://doi.org/10.1103/PhysRevX.10.021046} {\bibfield  {journal} {\bibinfo  {journal} {Phys. Rev. X}\ }\textbf {\bibinfo {volume} {10}},\ \bibinfo {pages} {021046} (\bibinfo {year} {2020})}\BibitemShut {NoStop}%
\bibitem [{\citenamefont {Machado}\ \emph {et~al.}(2020)\citenamefont {Machado}, \citenamefont {Else}, \citenamefont {Kahanamoku-Meyer}, \citenamefont {Nayak},\ and\ \citenamefont {Yao}}]{machadoLongRangePrethermalPhases2020}%
  \BibitemOpen
  \bibfield  {author} {\bibinfo {author} {\bibfnamefont {F.}~\bibnamefont {Machado}}, \bibinfo {author} {\bibfnamefont {D.~V.}\ \bibnamefont {Else}}, \bibinfo {author} {\bibfnamefont {G.~D.}\ \bibnamefont {Kahanamoku-Meyer}}, \bibinfo {author} {\bibfnamefont {C.}~\bibnamefont {Nayak}},\ and\ \bibinfo {author} {\bibfnamefont {N.~Y.}\ \bibnamefont {Yao}},\ }\bibfield  {title} {\bibinfo {title} {Long-{Range} {Prethermal} {Phases} of {Nonequilibrium} {Matter}},\ }\href {https://doi.org/10.1103/PhysRevX.10.011043} {\bibfield  {journal} {\bibinfo  {journal} {Phys. Rev. X}\ }\textbf {\bibinfo {volume} {10}},\ \bibinfo {pages} {011043} (\bibinfo {year} {2020})}\BibitemShut {NoStop}%
\bibitem [{\citenamefont {Pizzi}\ \emph {et~al.}(2021)\citenamefont {Pizzi}, \citenamefont {Knolle},\ and\ \citenamefont {Nunnenkamp}}]{pizziHigherorderFractionalDiscrete2021}%
  \BibitemOpen
  \bibfield  {author} {\bibinfo {author} {\bibfnamefont {A.}~\bibnamefont {Pizzi}}, \bibinfo {author} {\bibfnamefont {J.}~\bibnamefont {Knolle}},\ and\ \bibinfo {author} {\bibfnamefont {A.}~\bibnamefont {Nunnenkamp}},\ }\bibfield  {title} {\bibinfo {title} {Higher-order and fractional discrete time crystals in clean long-range interacting systems},\ }\href {https://doi.org/10.1038/s41467-021-22583-5} {\bibfield  {journal} {\bibinfo  {journal} {Nat. Commun.}\ }\textbf {\bibinfo {volume} {12}},\ \bibinfo {pages} {2341} (\bibinfo {year} {2021})}\BibitemShut {NoStop}%
\bibitem [{\citenamefont {Kyprianidis}\ \emph {et~al.}(2021)\citenamefont {Kyprianidis}, \citenamefont {Machado}, \citenamefont {Morong}, \citenamefont {Becker}, \citenamefont {Collins}, \citenamefont {Else}, \citenamefont {Feng}, \citenamefont {Hess}, \citenamefont {Nayak}, \citenamefont {Pagano}, \citenamefont {Yao},\ and\ \citenamefont {Monroe}}]{kyprianidisObservationPrethermalDiscrete2021}%
  \BibitemOpen
  \bibfield  {author} {\bibinfo {author} {\bibfnamefont {A.}~\bibnamefont {Kyprianidis}}, \bibinfo {author} {\bibfnamefont {F.}~\bibnamefont {Machado}}, \bibinfo {author} {\bibfnamefont {W.}~\bibnamefont {Morong}}, \bibinfo {author} {\bibfnamefont {P.}~\bibnamefont {Becker}}, \bibinfo {author} {\bibfnamefont {K.~S.}\ \bibnamefont {Collins}}, \bibinfo {author} {\bibfnamefont {D.~V.}\ \bibnamefont {Else}}, \bibinfo {author} {\bibfnamefont {L.}~\bibnamefont {Feng}}, \bibinfo {author} {\bibfnamefont {P.~W.}\ \bibnamefont {Hess}}, \bibinfo {author} {\bibfnamefont {C.}~\bibnamefont {Nayak}}, \bibinfo {author} {\bibfnamefont {G.}~\bibnamefont {Pagano}}, \bibinfo {author} {\bibfnamefont {N.~Y.}\ \bibnamefont {Yao}},\ and\ \bibinfo {author} {\bibfnamefont {C.}~\bibnamefont {Monroe}},\ }\bibfield  {title} {\bibinfo {title} {Observation of a prethermal discrete time crystal},\ }\href {https://doi.org/10.1126/science.abg8102} {\bibfield  {journal} {\bibinfo  {journal} {Science}\ }\textbf {\bibinfo {volume} {372}},\
  \bibinfo {pages} {1192} (\bibinfo {year} {2021})}\BibitemShut {NoStop}%
\bibitem [{\citenamefont {Santini}\ \emph {et~al.}(2022)\citenamefont {Santini}, \citenamefont {Santoro},\ and\ \citenamefont {Collura}}]{santiniCleanTwodimensionalFloquet2022}%
  \BibitemOpen
  \bibfield  {author} {\bibinfo {author} {\bibfnamefont {A.}~\bibnamefont {Santini}}, \bibinfo {author} {\bibfnamefont {G.~E.}\ \bibnamefont {Santoro}},\ and\ \bibinfo {author} {\bibfnamefont {M.}~\bibnamefont {Collura}},\ }\bibfield  {title} {\bibinfo {title} {Clean two-dimensional {{Floquet}} time crystal},\ }\href {https://doi.org/10.1103/PhysRevB.106.134301} {\bibfield  {journal} {\bibinfo  {journal} {Phys. Rev. B}\ }\textbf {\bibinfo {volume} {106}},\ \bibinfo {pages} {134301} (\bibinfo {year} {2022})}\BibitemShut {NoStop}%
\bibitem [{\citenamefont {Vu}\ and\ \citenamefont {Das~Sarma}(2023)}]{vuDissipativePrethermalDiscrete2023}%
  \BibitemOpen
  \bibfield  {author} {\bibinfo {author} {\bibfnamefont {D.}~\bibnamefont {Vu}}\ and\ \bibinfo {author} {\bibfnamefont {S.}~\bibnamefont {Das~Sarma}},\ }\bibfield  {title} {\bibinfo {title} {Dissipative {Prethermal} {Discrete} {Time} {Crystal}},\ }\href {https://doi.org/10.1103/PhysRevLett.130.130401} {\bibfield  {journal} {\bibinfo  {journal} {Phys. Rev. Lett.}\ }\textbf {\bibinfo {volume} {130}},\ \bibinfo {pages} {130401} (\bibinfo {year} {2023})}\BibitemShut {NoStop}%
\bibitem [{\citenamefont {Beatrez}\ \emph {et~al.}(2023)\citenamefont {Beatrez}, \citenamefont {Fleckenstein}, \citenamefont {Pillai}, \citenamefont {De~Leon~Sanchez}, \citenamefont {Akkiraju}, \citenamefont {Diaz~Alcala}, \citenamefont {Conti}, \citenamefont {Reshetikhin}, \citenamefont {Druga}, \citenamefont {Bukov},\ and\ \citenamefont {Ajoy}}]{beatrezCriticalPrethermalDiscrete2023}%
  \BibitemOpen
  \bibfield  {author} {\bibinfo {author} {\bibfnamefont {W.}~\bibnamefont {Beatrez}}, \bibinfo {author} {\bibfnamefont {C.}~\bibnamefont {Fleckenstein}}, \bibinfo {author} {\bibfnamefont {A.}~\bibnamefont {Pillai}}, \bibinfo {author} {\bibfnamefont {E.}~\bibnamefont {De~Leon~Sanchez}}, \bibinfo {author} {\bibfnamefont {A.}~\bibnamefont {Akkiraju}}, \bibinfo {author} {\bibfnamefont {J.}~\bibnamefont {Diaz~Alcala}}, \bibinfo {author} {\bibfnamefont {S.}~\bibnamefont {Conti}}, \bibinfo {author} {\bibfnamefont {P.}~\bibnamefont {Reshetikhin}}, \bibinfo {author} {\bibfnamefont {E.}~\bibnamefont {Druga}}, \bibinfo {author} {\bibfnamefont {M.}~\bibnamefont {Bukov}},\ and\ \bibinfo {author} {\bibfnamefont {A.}~\bibnamefont {Ajoy}},\ }\bibfield  {title} {\bibinfo {title} {Critical prethermal discrete time crystal created by two-frequency driving},\ }\href {https://doi.org/10.1038/s41567-022-01891-7} {\bibfield  {journal} {\bibinfo  {journal} {Nat. Phys.}\ }\textbf {\bibinfo {volume} {19}},\ \bibinfo {pages} {407}
  (\bibinfo {year} {2023})}\BibitemShut {NoStop}%
\bibitem [{\citenamefont {Xiang}\ \emph {et~al.}(2024)\citenamefont {Xiang} \emph {et~al.}}]{xiangLonglivedTopologicalTimecrystalline2024}%
  \BibitemOpen
  \bibfield  {author} {\bibinfo {author} {\bibfnamefont {L.}~\bibnamefont {Xiang}} \emph {et~al.},\ }\bibfield  {title} {\bibinfo {title} {Long-lived topological time-crystalline order on a quantum processor},\ }\href {https://doi.org/10.1038/s41467-024-53077-9} {\bibfield  {journal} {\bibinfo  {journal} {Nat. Commun.}\ }\textbf {\bibinfo {volume} {15}},\ \bibinfo {pages} {8963} (\bibinfo {year} {2024})}\BibitemShut {NoStop}%
\bibitem [{\citenamefont {Shinjo}\ \emph {et~al.}(2024)\citenamefont {Shinjo}, \citenamefont {Seki}, \citenamefont {Shirakawa}, \citenamefont {Sun},\ and\ \citenamefont {Yunoki}}]{shinjoUnveilingCleanTwodimensional2024}%
  \BibitemOpen
  \bibfield  {author} {\bibinfo {author} {\bibfnamefont {K.}~\bibnamefont {Shinjo}}, \bibinfo {author} {\bibfnamefont {K.}~\bibnamefont {Seki}}, \bibinfo {author} {\bibfnamefont {T.}~\bibnamefont {Shirakawa}}, \bibinfo {author} {\bibfnamefont {R.-Y.}\ \bibnamefont {Sun}},\ and\ \bibinfo {author} {\bibfnamefont {S.}~\bibnamefont {Yunoki}},\ }\href@noop {} {\bibinfo {title} {Unveiling clean two-dimensional discrete time quasicrystals on a digital quantum computer}} (\bibinfo {year} {2024}),\ \Eprint {https://arxiv.org/abs/2403.16718} {arXiv:2403.16718} \BibitemShut {NoStop}%
\bibitem [{\citenamefont {Jiang}\ \emph {et~al.}(2024)\citenamefont {Jiang}, \citenamefont {Yuan}, \citenamefont {Jiang}, \citenamefont {Deng},\ and\ \citenamefont {Machado}}]{jiangPrethermalTimeCrystallineCorner2024}%
  \BibitemOpen
  \bibfield  {author} {\bibinfo {author} {\bibfnamefont {S.}~\bibnamefont {Jiang}}, \bibinfo {author} {\bibfnamefont {D.}~\bibnamefont {Yuan}}, \bibinfo {author} {\bibfnamefont {W.}~\bibnamefont {Jiang}}, \bibinfo {author} {\bibfnamefont {D.-L.}\ \bibnamefont {Deng}},\ and\ \bibinfo {author} {\bibfnamefont {F.}~\bibnamefont {Machado}},\ }\href@noop {} {\bibinfo {title} {Prethermal {Time}-{Crystalline} {Corner} {Modes}}} (\bibinfo {year} {2024}),\ \Eprint {https://arxiv.org/abs/2406.01686} {arXiv:2406.01686} \BibitemShut {NoStop}%
\bibitem [{\citenamefont {Moon}\ \emph {et~al.}(2024)\citenamefont {Moon}, \citenamefont {Schindler}, \citenamefont {Smith}, \citenamefont {Druga}, \citenamefont {Zhang}, \citenamefont {Bukov},\ and\ \citenamefont {Ajoy}}]{moonDiscreteTimeCrystal2024}%
  \BibitemOpen
  \bibfield  {author} {\bibinfo {author} {\bibfnamefont {L.~J.~I.}\ \bibnamefont {Moon}}, \bibinfo {author} {\bibfnamefont {P.~M.}\ \bibnamefont {Schindler}}, \bibinfo {author} {\bibfnamefont {R.~J.}\ \bibnamefont {Smith}}, \bibinfo {author} {\bibfnamefont {E.}~\bibnamefont {Druga}}, \bibinfo {author} {\bibfnamefont {Z.-R.}\ \bibnamefont {Zhang}}, \bibinfo {author} {\bibfnamefont {M.}~\bibnamefont {Bukov}},\ and\ \bibinfo {author} {\bibfnamefont {A.}~\bibnamefont {Ajoy}},\ }\href@noop {} {\bibinfo {title} {Discrete {Time} {Crystal} {Sensing}}} (\bibinfo {year} {2024}),\ \Eprint {https://arxiv.org/abs/2410.05625} {arXiv:2410.05625} \BibitemShut {NoStop}%
\bibitem [{\citenamefont {Shukla}\ \emph {et~al.}(2025)\citenamefont {Shukla}, \citenamefont {Chotorlishvili}, \citenamefont {Mishra},\ and\ \citenamefont {Iemini}}]{shuklaPrethermalFloquetTime2024}%
  \BibitemOpen
  \bibfield  {author} {\bibinfo {author} {\bibfnamefont {R.~K.}\ \bibnamefont {Shukla}}, \bibinfo {author} {\bibfnamefont {L.}~\bibnamefont {Chotorlishvili}}, \bibinfo {author} {\bibfnamefont {S.~K.}\ \bibnamefont {Mishra}},\ and\ \bibinfo {author} {\bibfnamefont {F.}~\bibnamefont {Iemini}},\ }\bibfield  {title} {\bibinfo {title} {Prethermal {Floquet} time crystals in chiral multiferroic chains and applications as quantum sensors of {AC} fields},\ }\href {https://doi.org/10.1103/PhysRevB.111.024315} {\bibfield  {journal} {\bibinfo  {journal} {Phys. Rev. B}\ }\textbf {\bibinfo {volume} {111}},\ \bibinfo {pages} {024315} (\bibinfo {year} {2025})}\BibitemShut {NoStop}%
\bibitem [{\citenamefont {Ikeda}\ \emph {et~al.}(2024)\citenamefont {Ikeda}, \citenamefont {Sugiura},\ and\ \citenamefont {Polkovnikov}}]{ikedaRobustEffectiveGround2024}%
  \BibitemOpen
  \bibfield  {author} {\bibinfo {author} {\bibfnamefont {T.~N.}\ \bibnamefont {Ikeda}}, \bibinfo {author} {\bibfnamefont {S.}~\bibnamefont {Sugiura}},\ and\ \bibinfo {author} {\bibfnamefont {A.}~\bibnamefont {Polkovnikov}},\ }\bibfield  {title} {\bibinfo {title} {Robust {Effective} {Ground} {State} in a {Nonintegrable} {Floquet} {Quantum} {Circuit}},\ }\href {https://doi.org/10.1103/PhysRevLett.133.030401} {\bibfield  {journal} {\bibinfo  {journal} {Phys. Rev. Lett.}\ }\textbf {\bibinfo {volume} {133}},\ \bibinfo {pages} {030401} (\bibinfo {year} {2024})}\BibitemShut {NoStop}%
\bibitem [{\citenamefont {Wilson}(1974)}]{wilsonConfinementQuarks1974}%
  \BibitemOpen
  \bibfield  {author} {\bibinfo {author} {\bibfnamefont {K.~G.}\ \bibnamefont {Wilson}},\ }\bibfield  {title} {\bibinfo {title} {Confinement of quarks},\ }\href {https://doi.org/10.1103/PhysRevD.10.2445} {\bibfield  {journal} {\bibinfo  {journal} {Phys. Rev. D}\ }\textbf {\bibinfo {volume} {10}},\ \bibinfo {pages} {2445} (\bibinfo {year} {1974})}\BibitemShut {NoStop}%
\bibitem [{\citenamefont {Sulejmanpasic}\ \emph {et~al.}(2017)\citenamefont {Sulejmanpasic}, \citenamefont {Shao}, \citenamefont {Sandvik},\ and\ \citenamefont {{\"U}nsal}}]{sulejmanpasicConfinementBulkDeconfinement2017}%
  \BibitemOpen
  \bibfield  {author} {\bibinfo {author} {\bibfnamefont {T.}~\bibnamefont {Sulejmanpasic}}, \bibinfo {author} {\bibfnamefont {H.}~\bibnamefont {Shao}}, \bibinfo {author} {\bibfnamefont {A.~W.}\ \bibnamefont {Sandvik}},\ and\ \bibinfo {author} {\bibfnamefont {M.}~\bibnamefont {{\"U}nsal}},\ }\bibfield  {title} {\bibinfo {title} {Confinement in the {{Bulk}}, {{Deconfinement}} on the {{Wall}}: {{Infrared Equivalence}} between {{Compactified QCD}} and {{Quantum Magnets}}},\ }\href {https://doi.org/10.1103/PhysRevLett.119.091601} {\bibfield  {journal} {\bibinfo  {journal} {Phys. Rev. Lett.}\ }\textbf {\bibinfo {volume} {119}},\ \bibinfo {pages} {091601} (\bibinfo {year} {2017})}\BibitemShut {NoStop}%
\bibitem [{\citenamefont {Lake}\ \emph {et~al.}(2010)\citenamefont {Lake}, \citenamefont {Tsvelik}, \citenamefont {Notbohm}, \citenamefont {Alan~Tennant}, \citenamefont {Perring}, \citenamefont {Reehuis}, \citenamefont {Sekar}, \citenamefont {Krabbes},\ and\ \citenamefont {B{\"u}chner}}]{lakeConfinementFractionalQuantum2010}%
  \BibitemOpen
  \bibfield  {author} {\bibinfo {author} {\bibfnamefont {B.}~\bibnamefont {Lake}}, \bibinfo {author} {\bibfnamefont {A.~M.}\ \bibnamefont {Tsvelik}}, \bibinfo {author} {\bibfnamefont {S.}~\bibnamefont {Notbohm}}, \bibinfo {author} {\bibfnamefont {D.}~\bibnamefont {Alan~Tennant}}, \bibinfo {author} {\bibfnamefont {T.~G.}\ \bibnamefont {Perring}}, \bibinfo {author} {\bibfnamefont {M.}~\bibnamefont {Reehuis}}, \bibinfo {author} {\bibfnamefont {C.}~\bibnamefont {Sekar}}, \bibinfo {author} {\bibfnamefont {G.}~\bibnamefont {Krabbes}},\ and\ \bibinfo {author} {\bibfnamefont {B.}~\bibnamefont {B{\"u}chner}},\ }\bibfield  {title} {\bibinfo {title} {Confinement of fractional quantum number particles in a condensed-matter system},\ }\href {https://doi.org/10.1038/nphys1462} {\bibfield  {journal} {\bibinfo  {journal} {Nat. Phys.}\ }\textbf {\bibinfo {volume} {6}},\ \bibinfo {pages} {50} (\bibinfo {year} {2010})}\BibitemShut {NoStop}%
\bibitem [{\citenamefont {James}\ \emph {et~al.}(2019)\citenamefont {James}, \citenamefont {Konik},\ and\ \citenamefont {Robinson}}]{jamesNonthermalStatesArising2019}%
  \BibitemOpen
  \bibfield  {author} {\bibinfo {author} {\bibfnamefont {A.~J.~A.}\ \bibnamefont {James}}, \bibinfo {author} {\bibfnamefont {R.~M.}\ \bibnamefont {Konik}},\ and\ \bibinfo {author} {\bibfnamefont {N.~J.}\ \bibnamefont {Robinson}},\ }\bibfield  {title} {\bibinfo {title} {Nonthermal {{States Arising}} from {{Confinement}} in {{One}} and {{Two Dimensions}}},\ }\href {https://doi.org/10.1103/PhysRevLett.122.130603} {\bibfield  {journal} {\bibinfo  {journal} {Phys. Rev. Lett.}\ }\textbf {\bibinfo {volume} {122}},\ \bibinfo {pages} {130603} (\bibinfo {year} {2019})}\BibitemShut {NoStop}%
\bibitem [{\citenamefont {Ramos}\ \emph {et~al.}(2020)\citenamefont {Ramos}, \citenamefont {Lencs{\'e}s}, \citenamefont {Xavier},\ and\ \citenamefont {Pereira}}]{ramosConfinementBoundStates2020}%
  \BibitemOpen
  \bibfield  {author} {\bibinfo {author} {\bibfnamefont {F.~B.}\ \bibnamefont {Ramos}}, \bibinfo {author} {\bibfnamefont {M.}~\bibnamefont {Lencs{\'e}s}}, \bibinfo {author} {\bibfnamefont {J.~C.}\ \bibnamefont {Xavier}},\ and\ \bibinfo {author} {\bibfnamefont {R.~G.}\ \bibnamefont {Pereira}},\ }\bibfield  {title} {\bibinfo {title} {Confinement and bound states of bound states in a transverse-field two-leg {{Ising}} ladder},\ }\href {https://doi.org/10.1103/PhysRevB.102.014426} {\bibfield  {journal} {\bibinfo  {journal} {Phys. Rev. B}\ }\textbf {\bibinfo {volume} {102}},\ \bibinfo {pages} {014426} (\bibinfo {year} {2020})}\BibitemShut {NoStop}%
\bibitem [{\citenamefont {Kormos}\ \emph {et~al.}(2017)\citenamefont {Kormos}, \citenamefont {Collura}, \citenamefont {Tak{\'a}cs},\ and\ \citenamefont {Calabrese}}]{kormosRealtimeConfinementFollowing2017}%
  \BibitemOpen
  \bibfield  {author} {\bibinfo {author} {\bibfnamefont {M.}~\bibnamefont {Kormos}}, \bibinfo {author} {\bibfnamefont {M.}~\bibnamefont {Collura}}, \bibinfo {author} {\bibfnamefont {G.}~\bibnamefont {Tak{\'a}cs}},\ and\ \bibinfo {author} {\bibfnamefont {P.}~\bibnamefont {Calabrese}},\ }\bibfield  {title} {\bibinfo {title} {Real-time confinement following a quantum quench to a non-integrable model},\ }\href {https://doi.org/10.1038/nphys3934} {\bibfield  {journal} {\bibinfo  {journal} {Nat. Phys.}\ }\textbf {\bibinfo {volume} {13}},\ \bibinfo {pages} {246} (\bibinfo {year} {2017})}\BibitemShut {NoStop}%
\bibitem [{\citenamefont {Birnkammer}\ \emph {et~al.}(2022)\citenamefont {Birnkammer}, \citenamefont {Bastianello},\ and\ \citenamefont {Knap}}]{birnkammerPrethermalizationOnedimensionalQuantum2022}%
  \BibitemOpen
  \bibfield  {author} {\bibinfo {author} {\bibfnamefont {S.}~\bibnamefont {Birnkammer}}, \bibinfo {author} {\bibfnamefont {A.}~\bibnamefont {Bastianello}},\ and\ \bibinfo {author} {\bibfnamefont {M.}~\bibnamefont {Knap}},\ }\bibfield  {title} {\bibinfo {title} {Prethermalization in one-dimensional quantum many-body systems with confinement},\ }\href {https://doi.org/10.1038/s41467-022-35301-6} {\bibfield  {journal} {\bibinfo  {journal} {Nat. Commun.}\ }\textbf {\bibinfo {volume} {13}},\ \bibinfo {pages} {7663} (\bibinfo {year} {2022})}\BibitemShut {NoStop}%
\bibitem [{\citenamefont {Tindall}\ and\ \citenamefont {Sels}(2024)}]{tindallConfinementTransverseField2024}%
  \BibitemOpen
  \bibfield  {author} {\bibinfo {author} {\bibfnamefont {J.}~\bibnamefont {Tindall}}\ and\ \bibinfo {author} {\bibfnamefont {D.}~\bibnamefont {Sels}},\ }\bibfield  {title} {\bibinfo {title} {Confinement in the transverse field ising model on the heavy hex lattice},\ }\href {https://doi.org/10.1103/PhysRevLett.133.180402} {\bibfield  {journal} {\bibinfo  {journal} {Phys. Rev. Lett.}\ }\textbf {\bibinfo {volume} {133}},\ \bibinfo {pages} {180402} (\bibinfo {year} {2024})}\BibitemShut {NoStop}%
\bibitem [{\citenamefont {Collura}\ \emph {et~al.}(2022)\citenamefont {Collura}, \citenamefont {De~Luca}, \citenamefont {Rossini},\ and\ \citenamefont {Lerose}}]{colluraDiscreteTimeCrystallineResponse2022}%
  \BibitemOpen
  \bibfield  {author} {\bibinfo {author} {\bibfnamefont {M.}~\bibnamefont {Collura}}, \bibinfo {author} {\bibfnamefont {A.}~\bibnamefont {De~Luca}}, \bibinfo {author} {\bibfnamefont {D.}~\bibnamefont {Rossini}},\ and\ \bibinfo {author} {\bibfnamefont {A.}~\bibnamefont {Lerose}},\ }\bibfield  {title} {\bibinfo {title} {Discrete {{Time-Crystalline Response Stabilized}} by {{Domain-Wall Confinement}}},\ }\href {https://doi.org/10.1103/PhysRevX.12.031037} {\bibfield  {journal} {\bibinfo  {journal} {Phys. Rev. X}\ }\textbf {\bibinfo {volume} {12}},\ \bibinfo {pages} {031037} (\bibinfo {year} {2022})}\BibitemShut {NoStop}%
\bibitem [{\citenamefont {Pave{\v s}i{\'c}}\ \emph {et~al.}(2024)\citenamefont {Pave{\v s}i{\'c}}, \citenamefont {Jaschke},\ and\ \citenamefont {Montangero}}]{pavesicConstrainedDynamicsConfinement2024}%
  \BibitemOpen
  \bibfield  {author} {\bibinfo {author} {\bibfnamefont {L.}~\bibnamefont {Pave{\v s}i{\'c}}}, \bibinfo {author} {\bibfnamefont {D.}~\bibnamefont {Jaschke}},\ and\ \bibinfo {author} {\bibfnamefont {S.}~\bibnamefont {Montangero}},\ }\href@noop {} {\bibinfo {title} {Constrained dynamics and confinement in the two-dimensional quantum {{Ising}} model}} (\bibinfo {year} {2024}),\ \Eprint {https://arxiv.org/abs/2406.11979} {arXiv:2406.11979} \BibitemShut {NoStop}%
\bibitem [{\citenamefont {Tindall}\ \emph {et~al.}(2024)\citenamefont {Tindall}, \citenamefont {Fishman}, \citenamefont {Stoudenmire},\ and\ \citenamefont {Sels}}]{tindallEfficientTensorNetwork2024}%
  \BibitemOpen
  \bibfield  {author} {\bibinfo {author} {\bibfnamefont {J.}~\bibnamefont {Tindall}}, \bibinfo {author} {\bibfnamefont {M.}~\bibnamefont {Fishman}}, \bibinfo {author} {\bibfnamefont {E.~M.}\ \bibnamefont {Stoudenmire}},\ and\ \bibinfo {author} {\bibfnamefont {D.}~\bibnamefont {Sels}},\ }\bibfield  {title} {\bibinfo {title} {Efficient {{Tensor Network Simulation}} of {{IBM}}'s {{Eagle Kicked Ising Experiment}}},\ }\href {https://doi.org/10.1103/PRXQuantum.5.010308} {\bibfield  {journal} {\bibinfo  {journal} {PRX Quantum}\ }\textbf {\bibinfo {volume} {5}},\ \bibinfo {pages} {010308} (\bibinfo {year} {2024})}\BibitemShut {NoStop}%
\bibitem [{\citenamefont {Tindall}\ and\ \citenamefont {Fishman}(2023)}]{tindallGaugingTensorNetworks2023}%
  \BibitemOpen
  \bibfield  {author} {\bibinfo {author} {\bibfnamefont {J.}~\bibnamefont {Tindall}}\ and\ \bibinfo {author} {\bibfnamefont {M.}~\bibnamefont {Fishman}},\ }\bibfield  {title} {\bibinfo {title} {Gauging tensor networks with belief propagation},\ }\href {https://doi.org/10.21468/SciPostPhys.15.6.222} {\bibfield  {journal} {\bibinfo  {journal} {SciPost Phys.}\ }\textbf {\bibinfo {volume} {15}},\ \bibinfo {pages} {222} (\bibinfo {year} {2023})}\BibitemShut {NoStop}%
\bibitem [{\citenamefont {Bollob{\'a}s}(1998)}]{bollobasModernGraphTheory1998}%
  \BibitemOpen
  \bibfield  {author} {\bibinfo {author} {\bibfnamefont {B.}~\bibnamefont {Bollob{\'a}s}},\ }\href@noop {} {\emph {\bibinfo {title} {Modern Graph Theory}}},\ \bibinfo {series} {Graduate Texts in Mathematics}\ No.\ \bibinfo {number} {184}\ (\bibinfo  {publisher} {Springer},\ \bibinfo {address} {New York},\ \bibinfo {year} {1998})\BibitemShut {NoStop}%
\bibitem [{\citenamefont {Leifer}\ and\ \citenamefont {Poulin}(2008)}]{leiferQuantumGraphicalModels2008}%
  \BibitemOpen
  \bibfield  {author} {\bibinfo {author} {\bibfnamefont {M.}~\bibnamefont {Leifer}}\ and\ \bibinfo {author} {\bibfnamefont {D.}~\bibnamefont {Poulin}},\ }\bibfield  {title} {\bibinfo {title} {Quantum {Graphical} {Models} and {Belief} {Propagation}},\ }\href {https://doi.org/10.1016/j.aop.2007.10.001} {\bibfield  {journal} {\bibinfo  {journal} {Ann. Phys.}\ }\textbf {\bibinfo {volume} {323}},\ \bibinfo {pages} {1899} (\bibinfo {year} {2008})}\BibitemShut {NoStop}%
\bibitem [{\citenamefont {Alkabetz}\ and\ \citenamefont {Arad}(2021)}]{alkabetzTensorNetworksContraction2021}%
  \BibitemOpen
  \bibfield  {author} {\bibinfo {author} {\bibfnamefont {R.}~\bibnamefont {Alkabetz}}\ and\ \bibinfo {author} {\bibfnamefont {I.}~\bibnamefont {Arad}},\ }\bibfield  {title} {\bibinfo {title} {Tensor networks contraction and the belief propagation algorithm},\ }\href {https://doi.org/10.1103/PhysRevResearch.3.023073} {\bibfield  {journal} {\bibinfo  {journal} {Phys. Rev. Res.}\ }\textbf {\bibinfo {volume} {3}},\ \bibinfo {pages} {023073} (\bibinfo {year} {2021})}\BibitemShut {NoStop}%
\bibitem [{\citenamefont {Sahu}\ and\ \citenamefont {Swingle}(2022)}]{sahuEfficientTensorNetwork2022}%
  \BibitemOpen
  \bibfield  {author} {\bibinfo {author} {\bibfnamefont {S.}~\bibnamefont {Sahu}}\ and\ \bibinfo {author} {\bibfnamefont {B.}~\bibnamefont {Swingle}},\ }\href@noop {} {\bibinfo {title} {Efficient tensor network simulation of quantum many-body physics on sparse graphs}} (\bibinfo {year} {2022}),\ \Eprint {https://arxiv.org/abs/2206.04701} {arXiv:2206.04701} \BibitemShut {NoStop}%
\bibitem [{\citenamefont {Guo}\ \emph {et~al.}(2023)\citenamefont {Guo}, \citenamefont {Poletti},\ and\ \citenamefont {Arad}}]{guoBlockBeliefPropagation2023}%
  \BibitemOpen
  \bibfield  {author} {\bibinfo {author} {\bibfnamefont {C.}~\bibnamefont {Guo}}, \bibinfo {author} {\bibfnamefont {D.}~\bibnamefont {Poletti}},\ and\ \bibinfo {author} {\bibfnamefont {I.}~\bibnamefont {Arad}},\ }\bibfield  {title} {\bibinfo {title} {Block belief propagation algorithm for two-dimensional tensor networks},\ }\href {https://doi.org/10.1103/PhysRevB.108.125111} {\bibfield  {journal} {\bibinfo  {journal} {Phys. Rev. B}\ }\textbf {\bibinfo {volume} {108}},\ \bibinfo {pages} {125111} (\bibinfo {year} {2023})}\BibitemShut {NoStop}%
\bibitem [{\citenamefont {Else}\ \emph {et~al.}(2020{\natexlab{b}})\citenamefont {Else}, \citenamefont {Ho},\ and\ \citenamefont {Dumitrescu}}]{elseLongLivedInteractingPhases2020}%
  \BibitemOpen
  \bibfield  {author} {\bibinfo {author} {\bibfnamefont {D.~V.}\ \bibnamefont {Else}}, \bibinfo {author} {\bibfnamefont {W.~W.}\ \bibnamefont {Ho}},\ and\ \bibinfo {author} {\bibfnamefont {P.~T.}\ \bibnamefont {Dumitrescu}},\ }\bibfield  {title} {\bibinfo {title} {Long-{Lived} {Interacting} {Phases} of {Matter} {Protected} by {Multiple} {Time}-{Translation} {Symmetries} in {Quasiperiodically} {Driven} {Systems}},\ }\href {https://doi.org/10.1103/PhysRevX.10.021032} {\bibfield  {journal} {\bibinfo  {journal} {Phys. Rev. X}\ }\textbf {\bibinfo {volume} {10}},\ \bibinfo {pages} {021032} (\bibinfo {year} {2020}{\natexlab{b}})}\BibitemShut {NoStop}%
\bibitem [{\citenamefont {Autti}\ \emph {et~al.}(2018)\citenamefont {Autti}, \citenamefont {Eltsov},\ and\ \citenamefont {Volovik}}]{auttiObservationTimeQuasicrystal2018}%
  \BibitemOpen
  \bibfield  {author} {\bibinfo {author} {\bibfnamefont {S.}~\bibnamefont {Autti}}, \bibinfo {author} {\bibfnamefont {V.}~\bibnamefont {Eltsov}},\ and\ \bibinfo {author} {\bibfnamefont {G.}~\bibnamefont {Volovik}},\ }\bibfield  {title} {\bibinfo {title} {Observation of a {Time} {Quasicrystal} and {Its} {Transition} to a {Superfluid} {Time} {Crystal}},\ }\href {https://doi.org/10.1103/PhysRevLett.120.215301} {\bibfield  {journal} {\bibinfo  {journal} {Phys. Rev. Lett.}\ }\textbf {\bibinfo {volume} {120}},\ \bibinfo {pages} {215301} (\bibinfo {year} {2018})}\BibitemShut {NoStop}%
\bibitem [{\citenamefont {He}\ \emph {et~al.}(2024)\citenamefont {He}, \citenamefont {Ye}, \citenamefont {Gong}, \citenamefont {Yao}, \citenamefont {Liu}, \citenamefont {Murch}, \citenamefont {Yao},\ and\ \citenamefont {Zu}}]{heExperimentalRealizationDiscrete2024}%
  \BibitemOpen
  \bibfield  {author} {\bibinfo {author} {\bibfnamefont {G.}~\bibnamefont {He}}, \bibinfo {author} {\bibfnamefont {B.}~\bibnamefont {Ye}}, \bibinfo {author} {\bibfnamefont {R.}~\bibnamefont {Gong}}, \bibinfo {author} {\bibfnamefont {C.}~\bibnamefont {Yao}}, \bibinfo {author} {\bibfnamefont {Z.}~\bibnamefont {Liu}}, \bibinfo {author} {\bibfnamefont {K.~W.}\ \bibnamefont {Murch}}, \bibinfo {author} {\bibfnamefont {N.~Y.}\ \bibnamefont {Yao}},\ and\ \bibinfo {author} {\bibfnamefont {C.}~\bibnamefont {Zu}},\ }\href@noop {} {\bibinfo {title} {Experimental {{Realization}} of {{Discrete Time Quasi-Crystals}}}} (\bibinfo {year} {2024}),\ \Eprint {https://arxiv.org/abs/2403.17842} {arXiv:2403.17842} \BibitemShut {NoStop}%
\bibitem [{Note1()}]{Note1}%
  \BibitemOpen
  \bibinfo {note} {See Supplemental Material for a convergence analysis of the obtained BP-iTNS results, additional details on the finite entanglement scaling and a comparison to exact state vector simulations of Floquet dynamics in finite graphs.}\BibitemShut {Stop}%
\bibitem [{\citenamefont {Iemini}\ \emph {et~al.}(2018)\citenamefont {Iemini}, \citenamefont {Russomanno}, \citenamefont {Keeling}, \citenamefont {Schirò}, \citenamefont {Dalmonte},\ and\ \citenamefont {Fazio}}]{ieminiBoundaryTimeCrystals2018}%
  \BibitemOpen
  \bibfield  {author} {\bibinfo {author} {\bibfnamefont {F.}~\bibnamefont {Iemini}}, \bibinfo {author} {\bibfnamefont {A.}~\bibnamefont {Russomanno}}, \bibinfo {author} {\bibfnamefont {J.}~\bibnamefont {Keeling}}, \bibinfo {author} {\bibfnamefont {M.}~\bibnamefont {Schirò}}, \bibinfo {author} {\bibfnamefont {M.}~\bibnamefont {Dalmonte}},\ and\ \bibinfo {author} {\bibfnamefont {R.}~\bibnamefont {Fazio}},\ }\bibfield  {title} {\bibinfo {title} {Boundary {Time} {Crystals}},\ }\href {https://doi.org/10.1103/PhysRevLett.121.035301} {\bibfield  {journal} {\bibinfo  {journal} {Phys. Rev. Lett.}\ }\textbf {\bibinfo {volume} {121}},\ \bibinfo {pages} {035301} (\bibinfo {year} {2018})}\BibitemShut {NoStop}%
\bibitem [{\citenamefont {Piccitto}\ \emph {et~al.}(2021)\citenamefont {Piccitto}, \citenamefont {Wauters}, \citenamefont {Nori},\ and\ \citenamefont {Shammah}}]{piccittoSymmetriesConservedQuantities2021}%
  \BibitemOpen
  \bibfield  {author} {\bibinfo {author} {\bibfnamefont {G.}~\bibnamefont {Piccitto}}, \bibinfo {author} {\bibfnamefont {M.}~\bibnamefont {Wauters}}, \bibinfo {author} {\bibfnamefont {F.}~\bibnamefont {Nori}},\ and\ \bibinfo {author} {\bibfnamefont {N.}~\bibnamefont {Shammah}},\ }\bibfield  {title} {\bibinfo {title} {Symmetries and conserved quantities of boundary time crystals in generalized spin models},\ }\href {https://doi.org/10.1103/PhysRevB.104.014307} {\bibfield  {journal} {\bibinfo  {journal} {Phys. Rev. B}\ }\textbf {\bibinfo {volume} {104}},\ \bibinfo {pages} {014307} (\bibinfo {year} {2021})}\BibitemShut {NoStop}%
\bibitem [{\citenamefont {Carollo}\ and\ \citenamefont {Lesanovsky}(2022)}]{carolloExactSolutionBoundary2022}%
  \BibitemOpen
  \bibfield  {author} {\bibinfo {author} {\bibfnamefont {F.}~\bibnamefont {Carollo}}\ and\ \bibinfo {author} {\bibfnamefont {I.}~\bibnamefont {Lesanovsky}},\ }\bibfield  {title} {\bibinfo {title} {Exact solution of a boundary time-crystal phase transition: {Time}-translation symmetry breaking and non-{Markovian} dynamics of correlations},\ }\href {https://doi.org/10.1103/PhysRevA.105.L040202} {\bibfield  {journal} {\bibinfo  {journal} {Phys. Rev. A}\ }\textbf {\bibinfo {volume} {105}},\ \bibinfo {pages} {L040202} (\bibinfo {year} {2022})}\BibitemShut {NoStop}%
\bibitem [{\citenamefont {Zhang}\ \emph {et~al.}(2022)\citenamefont {Zhang} \emph {et~al.}}]{zhangDigitalQuantumSimulation2022}%
  \BibitemOpen
  \bibfield  {author} {\bibinfo {author} {\bibfnamefont {X.}~\bibnamefont {Zhang}} \emph {et~al.},\ }\bibfield  {title} {\bibinfo {title} {Digital quantum simulation of {Floquet} symmetry-protected topological phases},\ }\href {https://doi.org/10.1038/s41586-022-04854-3} {\bibfield  {journal} {\bibinfo  {journal} {Nature}\ }\textbf {\bibinfo {volume} {607}},\ \bibinfo {pages} {468} (\bibinfo {year} {2022})}\BibitemShut {NoStop}%
\bibitem [{\citenamefont {Bhowmick}\ \emph {et~al.}(2023)\citenamefont {Bhowmick}, \citenamefont {Sun}, \citenamefont {Yang},\ and\ \citenamefont {Sengupta}}]{bhowmickDiscreteTimeCrystal2023}%
  \BibitemOpen
  \bibfield  {author} {\bibinfo {author} {\bibfnamefont {D.}~\bibnamefont {Bhowmick}}, \bibinfo {author} {\bibfnamefont {H.}~\bibnamefont {Sun}}, \bibinfo {author} {\bibfnamefont {B.}~\bibnamefont {Yang}},\ and\ \bibinfo {author} {\bibfnamefont {P.}~\bibnamefont {Sengupta}},\ }\bibfield  {title} {\bibinfo {title} {Discrete time crystal made of topological edge magnons},\ }\href {https://doi.org/10.1103/PhysRevB.108.014434} {\bibfield  {journal} {\bibinfo  {journal} {Phys. Rev. B}\ }\textbf {\bibinfo {volume} {108}},\ \bibinfo {pages} {014434} (\bibinfo {year} {2023})}\BibitemShut {NoStop}%
\bibitem [{\citenamefont {Kim}\ \emph {et~al.}(2023)\citenamefont {Kim} \emph {et~al.}}]{kimEvidenceUtilityQuantum2023}%
  \BibitemOpen
  \bibfield  {author} {\bibinfo {author} {\bibfnamefont {Y.}~\bibnamefont {Kim}} \emph {et~al.},\ }\bibfield  {title} {\bibinfo {title} {Evidence for the utility of quantum computing before fault tolerance},\ }\href {https://doi.org/10.1038/s41586-023-06096-3} {\bibfield  {journal} {\bibinfo  {journal} {Nature}\ }\textbf {\bibinfo {volume} {618}},\ \bibinfo {pages} {500} (\bibinfo {year} {2023})}\BibitemShut {NoStop}%
\bibitem [{\citenamefont {Arute}\ \emph {et~al.}(2019)\citenamefont {Arute} \emph {et~al.}}]{aruteQuantumSupremacyUsing2019}%
  \BibitemOpen
  \bibfield  {author} {\bibinfo {author} {\bibfnamefont {F.}~\bibnamefont {Arute}} \emph {et~al.},\ }\bibfield  {title} {\bibinfo {title} {Quantum supremacy using a programmable superconducting processor},\ }\href {https://doi.org/10.1038/s41586-019-1666-5} {\bibfield  {journal} {\bibinfo  {journal} {Nature}\ }\textbf {\bibinfo {volume} {574}},\ \bibinfo {pages} {505} (\bibinfo {year} {2019})}\BibitemShut {NoStop}%
\bibitem [{\citenamefont {Boothby}\ \emph {et~al.}(2020)\citenamefont {Boothby}, \citenamefont {Bunyk}, \citenamefont {Raymond},\ and\ \citenamefont {Roy}}]{boothbyNextGenerationTopologyDWave2020}%
  \BibitemOpen
  \bibfield  {author} {\bibinfo {author} {\bibfnamefont {K.}~\bibnamefont {Boothby}}, \bibinfo {author} {\bibfnamefont {P.}~\bibnamefont {Bunyk}}, \bibinfo {author} {\bibfnamefont {J.}~\bibnamefont {Raymond}},\ and\ \bibinfo {author} {\bibfnamefont {A.}~\bibnamefont {Roy}},\ }\href@noop {} {\bibinfo {title} {Next-{Generation} {Topology} of {D}-{Wave} {Quantum} {Processors}}} (\bibinfo {year} {2020}),\ \Eprint {https://arxiv.org/abs/2003.00133} {arXiv:2003.00133} \BibitemShut {NoStop}%
\bibitem [{\citenamefont {Mi}\ \emph {et~al.}(2022)\citenamefont {Mi} \emph {et~al.}}]{miTimecrystallineEigenstateOrder2022}%
  \BibitemOpen
  \bibfield  {author} {\bibinfo {author} {\bibfnamefont {X.}~\bibnamefont {Mi}} \emph {et~al.},\ }\bibfield  {title} {\bibinfo {title} {Time-crystalline eigenstate order on a quantum processor},\ }\href {https://doi.org/10.1038/s41586-021-04257-w} {\bibfield  {journal} {\bibinfo  {journal} {Nature}\ }\textbf {\bibinfo {volume} {601}},\ \bibinfo {pages} {531} (\bibinfo {year} {2022})}\BibitemShut {NoStop}%
\bibitem [{\citenamefont {King}\ \emph {et~al.}(2025)\citenamefont {King} \emph {et~al.}}]{kingComputationalSupremacyQuantum2024}%
  \BibitemOpen
  \bibfield  {author} {\bibinfo {author} {\bibfnamefont {A.~D.}\ \bibnamefont {King}} \emph {et~al.},\ }\bibfield  {title} {\bibinfo {title} {Beyond-classical computation in quantum simulation},\ }\href {https://doi.org/10.1126/science.ado6285} {\bibfield  {journal} {\bibinfo  {journal} {Science}\ ,\ \bibinfo {pages} {eado6285}} (\bibinfo {year} {2025})}\BibitemShut {NoStop}%
\bibitem [{\citenamefont {Taie}\ \emph {et~al.}(2015)\citenamefont {Taie}, \citenamefont {Ozawa}, \citenamefont {Ichinose}, \citenamefont {Nishio}, \citenamefont {Nakajima},\ and\ \citenamefont {Takahashi}}]{taieCoherentDrivingFreezing2015}%
  \BibitemOpen
  \bibfield  {author} {\bibinfo {author} {\bibfnamefont {S.}~\bibnamefont {Taie}}, \bibinfo {author} {\bibfnamefont {H.}~\bibnamefont {Ozawa}}, \bibinfo {author} {\bibfnamefont {T.}~\bibnamefont {Ichinose}}, \bibinfo {author} {\bibfnamefont {T.}~\bibnamefont {Nishio}}, \bibinfo {author} {\bibfnamefont {S.}~\bibnamefont {Nakajima}},\ and\ \bibinfo {author} {\bibfnamefont {Y.}~\bibnamefont {Takahashi}},\ }\bibfield  {title} {\bibinfo {title} {Coherent driving and freezing of bosonic matter wave in an optical {Lieb} lattice},\ }\href {https://doi.org/10.1126/sciadv.1500854} {\bibfield  {journal} {\bibinfo  {journal} {Sci. Adv.}\ }\textbf {\bibinfo {volume} {1}},\ \bibinfo {pages} {e1500854} (\bibinfo {year} {2015})}\BibitemShut {NoStop}%
\bibitem [{\citenamefont {Periwal}\ \emph {et~al.}(2021)\citenamefont {Periwal}, \citenamefont {Cooper}, \citenamefont {Kunkel}, \citenamefont {Wienand}, \citenamefont {Davis},\ and\ \citenamefont {Schleier-Smith}}]{periwalProgrammableInteractionsEmergent2021}%
  \BibitemOpen
  \bibfield  {author} {\bibinfo {author} {\bibfnamefont {A.}~\bibnamefont {Periwal}}, \bibinfo {author} {\bibfnamefont {E.~S.}\ \bibnamefont {Cooper}}, \bibinfo {author} {\bibfnamefont {P.}~\bibnamefont {Kunkel}}, \bibinfo {author} {\bibfnamefont {J.~F.}\ \bibnamefont {Wienand}}, \bibinfo {author} {\bibfnamefont {E.~J.}\ \bibnamefont {Davis}},\ and\ \bibinfo {author} {\bibfnamefont {M.}~\bibnamefont {Schleier-Smith}},\ }\bibfield  {title} {\bibinfo {title} {Programmable interactions and emergent geometry in an array of atom clouds},\ }\href {https://doi.org/10.1038/s41586-021-04156-0} {\bibfield  {journal} {\bibinfo  {journal} {Nature}\ }\textbf {\bibinfo {volume} {600}},\ \bibinfo {pages} {630} (\bibinfo {year} {2021})}\BibitemShut {NoStop}%
\bibitem [{\citenamefont {Lebrat}\ \emph {et~al.}(2024)\citenamefont {Lebrat}, \citenamefont {Kale}, \citenamefont {Kendrick}, \citenamefont {Xu}, \citenamefont {Gang}, \citenamefont {Nikolaenko}, \citenamefont {Sachdev},\ and\ \citenamefont {Greiner}}]{lebratFerrimagnetismUltracoldFermions2024}%
  \BibitemOpen
  \bibfield  {author} {\bibinfo {author} {\bibfnamefont {M.}~\bibnamefont {Lebrat}}, \bibinfo {author} {\bibfnamefont {A.}~\bibnamefont {Kale}}, \bibinfo {author} {\bibfnamefont {L.~H.}\ \bibnamefont {Kendrick}}, \bibinfo {author} {\bibfnamefont {M.}~\bibnamefont {Xu}}, \bibinfo {author} {\bibfnamefont {Y.}~\bibnamefont {Gang}}, \bibinfo {author} {\bibfnamefont {A.}~\bibnamefont {Nikolaenko}}, \bibinfo {author} {\bibfnamefont {S.}~\bibnamefont {Sachdev}},\ and\ \bibinfo {author} {\bibfnamefont {M.}~\bibnamefont {Greiner}},\ }\href@noop {} {\bibinfo {title} {Ferrimagnetism of ultracold fermions in a multi-band {Hubbard} system}} (\bibinfo {year} {2024}),\ \Eprint {https://arxiv.org/abs/2404.17555} {arXiv:2404.17555} \BibitemShut {NoStop}%
\bibitem [{\citenamefont {Bu\v{c}a}\ \emph {et~al.}(2019)\citenamefont {Bu\v{c}a}, \citenamefont {Tindall},\ and\ \citenamefont {Jaksch}}]{bucaNonstationaryCoherentQuantum2019}%
  \BibitemOpen
  \bibfield  {author} {\bibinfo {author} {\bibfnamefont {B.}~\bibnamefont {Bu\v{c}a}}, \bibinfo {author} {\bibfnamefont {J.}~\bibnamefont {Tindall}},\ and\ \bibinfo {author} {\bibfnamefont {D.}~\bibnamefont {Jaksch}},\ }\bibfield  {title} {\bibinfo {title} {Non-stationary coherent quantum many-body dynamics through dissipation},\ }\href {https://doi.org/10.1038/s41467-019-09757-y} {\bibfield  {journal} {\bibinfo  {journal} {Nat. Commun.}\ }\textbf {\bibinfo {volume} {10}},\ \bibinfo {pages} {1730} (\bibinfo {year} {2019})}\BibitemShut {NoStop}%
\bibitem [{\citenamefont {Kongkhambut}\ \emph {et~al.}(2022)\citenamefont {Kongkhambut}, \citenamefont {Skulte}, \citenamefont {Mathey}, \citenamefont {Cosme}, \citenamefont {Hemmerich},\ and\ \citenamefont {Keßler}}]{kongkhambutObservationContinuousTime2022}%
  \BibitemOpen
  \bibfield  {author} {\bibinfo {author} {\bibfnamefont {P.}~\bibnamefont {Kongkhambut}}, \bibinfo {author} {\bibfnamefont {J.}~\bibnamefont {Skulte}}, \bibinfo {author} {\bibfnamefont {L.}~\bibnamefont {Mathey}}, \bibinfo {author} {\bibfnamefont {J.~G.}\ \bibnamefont {Cosme}}, \bibinfo {author} {\bibfnamefont {A.}~\bibnamefont {Hemmerich}},\ and\ \bibinfo {author} {\bibfnamefont {H.}~\bibnamefont {Keßler}},\ }\bibfield  {title} {\bibinfo {title} {Observation of a continuous time crystal},\ }\href {https://doi.org/10.1126/science.abo3382} {\bibfield  {journal} {\bibinfo  {journal} {Science}\ }\textbf {\bibinfo {volume} {377}},\ \bibinfo {pages} {670} (\bibinfo {year} {2022})}\BibitemShut {NoStop}%
\bibitem [{\citenamefont {Carraro-Haddad}\ \emph {et~al.}(2024)\citenamefont {Carraro-Haddad}, \citenamefont {Chafatinos}, \citenamefont {Kuznetsov}, \citenamefont {Papuccio-Fernández}, \citenamefont {Reynoso}, \citenamefont {Bruchhausen}, \citenamefont {Biermann}, \citenamefont {Santos}, \citenamefont {Usaj},\ and\ \citenamefont {Fainstein}}]{carraro-haddadSolidstateContinuousTime2024}%
  \BibitemOpen
  \bibfield  {author} {\bibinfo {author} {\bibfnamefont {I.}~\bibnamefont {Carraro-Haddad}}, \bibinfo {author} {\bibfnamefont {D.~L.}\ \bibnamefont {Chafatinos}}, \bibinfo {author} {\bibfnamefont {A.~S.}\ \bibnamefont {Kuznetsov}}, \bibinfo {author} {\bibfnamefont {I.~A.}\ \bibnamefont {Papuccio-Fernández}}, \bibinfo {author} {\bibfnamefont {A.~A.}\ \bibnamefont {Reynoso}}, \bibinfo {author} {\bibfnamefont {A.}~\bibnamefont {Bruchhausen}}, \bibinfo {author} {\bibfnamefont {K.}~\bibnamefont {Biermann}}, \bibinfo {author} {\bibfnamefont {P.~V.}\ \bibnamefont {Santos}}, \bibinfo {author} {\bibfnamefont {G.}~\bibnamefont {Usaj}},\ and\ \bibinfo {author} {\bibfnamefont {A.}~\bibnamefont {Fainstein}},\ }\bibfield  {title} {\bibinfo {title} {Solid-state continuous time crystal in a polariton condensate with a built-in mechanical clock},\ }\href {https://doi.org/10.1126/science.adn7087} {\bibfield  {journal} {\bibinfo  {journal} {Science}\ }\textbf {\bibinfo {volume} {384}},\ \bibinfo {pages} {995} (\bibinfo {year}
  {2024})}\BibitemShut {NoStop}%
\bibitem [{\citenamefont {Seibold}\ \emph {et~al.}(2020)\citenamefont {Seibold}, \citenamefont {Rota},\ and\ \citenamefont {Savona}}]{seiboldDissipativeTimeCrystal2020}%
  \BibitemOpen
  \bibfield  {author} {\bibinfo {author} {\bibfnamefont {K.}~\bibnamefont {Seibold}}, \bibinfo {author} {\bibfnamefont {R.}~\bibnamefont {Rota}},\ and\ \bibinfo {author} {\bibfnamefont {V.}~\bibnamefont {Savona}},\ }\bibfield  {title} {\bibinfo {title} {Dissipative time crystal in an asymmetric nonlinear photonic dimer},\ }\href {https://doi.org/10.1103/PhysRevA.101.033839} {\bibfield  {journal} {\bibinfo  {journal} {Phys. Rev. A}\ }\textbf {\bibinfo {volume} {101}},\ \bibinfo {pages} {033839} (\bibinfo {year} {2020})}\BibitemShut {NoStop}%
\bibitem [{ITe(2024)}]{ITensorNetworksJl2024}%
  \BibitemOpen
  \href@noop {} {\bibinfo {title} {{{iTensorNetworks}}.jl}},\ \bibinfo {howpublished} {\url{https://github.com/mtfishman/ITensorNetworks.jl}} (\bibinfo {year} {2024})\BibitemShut {NoStop}%
\bibitem [{\citenamefont {Bezanson}\ \emph {et~al.}(2017)\citenamefont {Bezanson}, \citenamefont {Edelman}, \citenamefont {Karpinski},\ and\ \citenamefont {Shah}}]{bezansonJuliaFreshApproach2017}%
  \BibitemOpen
  \bibfield  {author} {\bibinfo {author} {\bibfnamefont {J.}~\bibnamefont {Bezanson}}, \bibinfo {author} {\bibfnamefont {A.}~\bibnamefont {Edelman}}, \bibinfo {author} {\bibfnamefont {S.}~\bibnamefont {Karpinski}},\ and\ \bibinfo {author} {\bibfnamefont {V.~B.}\ \bibnamefont {Shah}},\ }\bibfield  {title} {\bibinfo {title} {Julia: {{A Fresh Approach}} to {{Numerical Computing}}},\ }\href {https://doi.org/10.1137/141000671} {\bibfield  {journal} {\bibinfo  {journal} {SIAM Rev.}\ }\textbf {\bibinfo {volume} {59}},\ \bibinfo {pages} {65} (\bibinfo {year} {2017})}\BibitemShut {NoStop}%
\bibitem [{\citenamefont {Fishman}\ \emph {et~al.}(2022)\citenamefont {Fishman}, \citenamefont {White},\ and\ \citenamefont {Stoudenmire}}]{fishmanITensorSoftwareLibrary2022}%
  \BibitemOpen
  \bibfield  {author} {\bibinfo {author} {\bibfnamefont {M.}~\bibnamefont {Fishman}}, \bibinfo {author} {\bibfnamefont {S.}~\bibnamefont {White}},\ and\ \bibinfo {author} {\bibfnamefont {E.}~\bibnamefont {Stoudenmire}},\ }\bibfield  {title} {\bibinfo {title} {The {{ITensor Software Library}} for {{Tensor Network Calculations}}},\ }\href {https://doi.org/10.21468/SciPostPhysCodeb.4} {\bibfield  {journal} {\bibinfo  {journal} {SciPost Phys. Codebases}\ ,\ \bibinfo {pages} {4}} (\bibinfo {year} {2022})}\BibitemShut {NoStop}%
\end{thebibliography}%
